\documentclass[5p,authoryear,preprint,times]{elsarticle}

\usepackage{natbib}
\usepackage{amsmath}
\usepackage{amssymb}
\usepackage{enumitem}
\usepackage{graphicx}
\usepackage{siunitx}
\usepackage{chapterbib}
\setlist{noitemsep}

\usepackage{url}

\newcommand{\ud}{\mathrm{d}}

\newcommand{\ue}{\mathrm{e}}
\newcommand{\p}{\partial}
\newcommand{\We}{\mathrm{We}}
\newcommand{\ReNum}{\mathrm{Re}}

\newcommand{\Mo}{\mathrm{Mo}}
\newcommand{\Eo}{\mathrm{Eo}}

\begin{document}
\begin{frontmatter}

\title{Vertical mixing in oil spill modelling}
\author[1,2]{Tor Nordam\corref{cor}}
\author[1]{J\o{}rgen Skancke}
\author[3,4]{Rodrigo Duran}
\author[5]{Chris Barker}

\address[1]{SINTEF Ocean, Trondheim, Norway}
\address[2]{Norwegian University of Science and Technology, Trondheim, Norway}
\address[3]{National Energy Technology Laboratory, U.S. Department of Energy, USA}
\address[4]{Theiss Research, USA}
\address[5]{Office of Response and Restoration, Emergency Response Division, National Oceanic and Atmospheric Administration, USA}

\cortext[cor]{tor.nordam@sintef.no}

\begin{abstract}
    The main focus of marine oil spill modelling is often on \emph{where} the
    oil will end up, \emph{i.e.}, on the horizontal transport. However, due to current
    shear, wind drag, and the different physical, chemical and biological
    processes that affect oil differently on the surface and in the water
    column, modelling the vertical distribution of the oil is essential for
    modelling the horizontal transport.

    In this work, we review and present models for a number of physical
    processes that influence the vertical transport of oil, including wave
    entrainment, droplet rise, vertical turbulent mixing, and surfacing. We aim
    to provide enough detail for the reader to be able to understand and
    implement the models, and to provide references to further reading.
    Mathematical and numerical details are included, particularly on the
    advection and diffusion of particles. We also present and discuss some
    common numerical pitfalls that may be a bit subtle, but which can cause
    significant errors.
\end{abstract}

\end{frontmatter}

This chapter is intended to give a good overview of the processes which relate
to the vertical movement of oil spilled in the ocean, namely turbulent mixing,
buoyancy, and entrainment. These processes will be explained in terms of their
physics, approaches to numerical modelling, and through the historical
background on how the research field has developed.

The aim of this chapter is that the reader should be capable of formulating a
reasonable ``one-dimensional'' oil spill model.

 \section{Introduction}

An essential aspect of oil spill modelling is to capture the different processes
that influence fate and behavior of oil on the ocean surface and oil in the
water column. Surface oil is exposed to the atmosphere, wind, and waves, and
undergoes surface spreading, evaporation, emulsification, and entrainment due to
breaking waves. Submerged oil, on the other hand, experiences to a greater
degree dissolution, microbial biodegradation and three-dimensional dispersion.
In terms of interaction, surface oil may foul birds and surface-interacting
mammals, and may be washed ashore to cover coastline habitats. Subsurface oil in
dissolved form represents an exposure risk to marine life. This is especially
the case for early life stages like eggs and larvae. Oil in droplet form
may for example cause toxicity by adhering to the surface of fish
eggs \citep{hansen2018adhesion}.

In addition to the distinction between surface and submerged oil, the vertical
distribution of oil within the water column has a major impact on horizontal
transport, due to current shear. The importance of vertical mixing for
horizontal transport has been known for a long time. Both \citet{bowden1965} and
\citet{okubo1968} suggest that \citet{bowles1958} were the first to use the term
``shear effect'' in relation to mixing and transport in the sea, in a paper on
dilution of radioactive waste water. \citet{okubo1968} states that the shear
effect is ``the dispersion of a vertical column of fluid due to the variation of
velocity with depth combined with vertical diffusion''.

Surface wind can create strong vertical gradients of current speed and direction
in the upper meters of the water column \citep{Laxague2018,Fernandez1996}, thus
leading to dispersion of oil submerged at different depths. Vertical density
gradients can also act to separate vertical layers, allowing them to move in
different horizontal directions, again contributing to current shear.

In the context of oil spill modelling, \citet{johansen1982} provides an early
discussion of the importance of vertical distribution for determining horizontal
transport. He describes the continuous exchange of oil between the surface and
the subsurface due to breaking waves and surfacing, discussing the importance of
rise speed and the droplet size distribution produced by natural entrainment. He
also explicitly formulates a one-dimensional Eulerian model for the vertical
transport of oil droplets, based on the advection-diffusion equation, and uses
this model to discuss the implications for the drift and weathering of surface
oil.

Another early work treating an oil spill as a three-dimensional process is that
of~\citet{elliott_shear_1986}.  This paper describes the elongation of an oil
slick in the direction of the wind, attributing the effect to vertical current
shear, combined with the continuous exchange of oil between the surface and the
subsurface.  The downwards process is driven by turbulent mixing from waves, and
the upwards process by the buoyancy of the oil droplets.
\citet{elliott_shear_1986} also formulated a three-dimensional random-walk based
Lagrangian particle model for an oil slick. While that model allowed
sufficiently large oil droplets to remain at the surface, as their rise due to
buoyancy would always dominate the random displacement due to diffusion, it did
not include a slick formation process or a separate state for surface oil.

In modern oil spill models, surfaced oil is usually assumed to form continuous
patches, while submerged oil is in the form of individual droplets of different
sizes.  More advanced models also include a non-buoyant dissolved oil fraction.
The mass exchange between the surface and subsurface compartments is a function
of the state of the oil and the state of the wave field, the latter of which can
be parameterised from wind speed and fetch length, or modelled by a wave model
(coupled to the ocean model, or separate). To calculate entrainment, an oil
spill model must predict the mass of oil entrained per area and time, over what
depth that oil should be distributed, and the droplet size distribution of the
entrained oil. Surfacing of oil, on the other hand, is found as a balance
between the vertical rise of droplets and turbulent mixing. Vertical transport
brings droplets toward the surface, while turbulent mixing tends to distribute
oil droplets over a certain depth. Strong turbulent mixing will therefore reduce
the amount of oil surfacing by reducing the concentration of oil droplets in the
near-surface water layer.

%\subsection{Overview of the chapter contents}

%%%%%%%%%%%%%%%%%%%%%%%%%%%%%%%%%%%%%%
%%%% Vertical mixing in the ocean %%%%
%%%%%%%%%%%%%%%%%%%%%%%%%%%%%%%%%%%%%%

\section{Vertical mixing in the ocean} \label{sec:ocean}

In this section, we give a description of the mechanisms behind vertical
mixing in the ocean. Some oceanographic background is given, though for further
information the interested reader may refer to, \emph{e.g.}, \emph{Introduction to
ocean turbulence}~\citep{thorpe2007}, and the classic work \emph{Turbulent
diffusion in the environment} \citep[see particularly chapters III, V and
VI]{csanady1973}.

%Topics to cover: \begin{itemize} \item Mixed layer \item Stratified flow \item
%Origins of turbulence in the ocean: \begin{itemize} \item Waves \item Sea-bed
%friction \item Current shear \item Sea ice?  \end{itemize} \item Langmuir
%circulation \item Further description of sea ice?  \item Some remarks about
%difference in horizontal and vertical scales, role of stratification
%\end{itemize}

\subsection{Turbulent diffusion}

Molecular diffusion is a fundamental physical process, caused by the random
motion of molecules in gases and liquids, occurring even in completely stagnant
conditions. The effect of this random motion is to reduce gradients in
concentration, leading eventually to an even distribution of, \emph{e.g.}, dissolved
chemicals in water. The rate of molecular diffusion depends on temperature and
the relative size and properties of the molecules involved, but this is in all
cases a relatively slow process. As an example, \citet{lee2004} did experiments
on the diffusion of ink in water, and found that a droplet of ink took about one
minute to spread to a radius of 1 cm, in water at room temperature.

In contrast, we would expect a droplet of ink to be evenly distributed in a
glass of water within seconds, if the water was stirred. This process is often
called turbulent diffusion, or turbulent mixing, and is akin to what happens in
the ocean, where the origins of the turbulent mixing can for example be breaking
waves, current shear, bottom friction, overturning, etc.

Despite the name, turbulent diffusion is not a pure diffusion process, but
rather a combination of an advection process and molecular diffusion (see,
\emph{e.g.},
\citet[pp. 20--21]{thorpe2005} for a particularly clear description).  The
crucial point is that turbulence causes stirring at a wide range of spatial
scales, dramatically increasing the area of interface between regions of high
and low concentrations. Fick's law (see, \emph{e.g.}, \citet[p.  4]{csanady1973})
states that the diffusive flux of a substance (\emph{i.e.}, amount of substance
transported per area per time) is given by
\begin{align}
    \label{eq:fick}
    j_D = -K \frac{\ud}{\ud x} C(x),
\end{align}
where $K$ is the diffusion parameter, and $C$ is the concentration of a
substance. If we consider two initially separated volumes of water, with
different concentrations of some substance, then it is clear that mixing will be
faster if the area of the interface between the two volumes increases. This is
precisely what turbulent mixing achieves, and the effect can be quite dramatic,
increasing the effective mixing by many orders of magnitude (see, \emph{e.g.},
\citet[pp. 8-10]{tennekes1972}).

\begin{figure*}
    \begin{center}
    \includegraphics[width=0.32\textwidth]{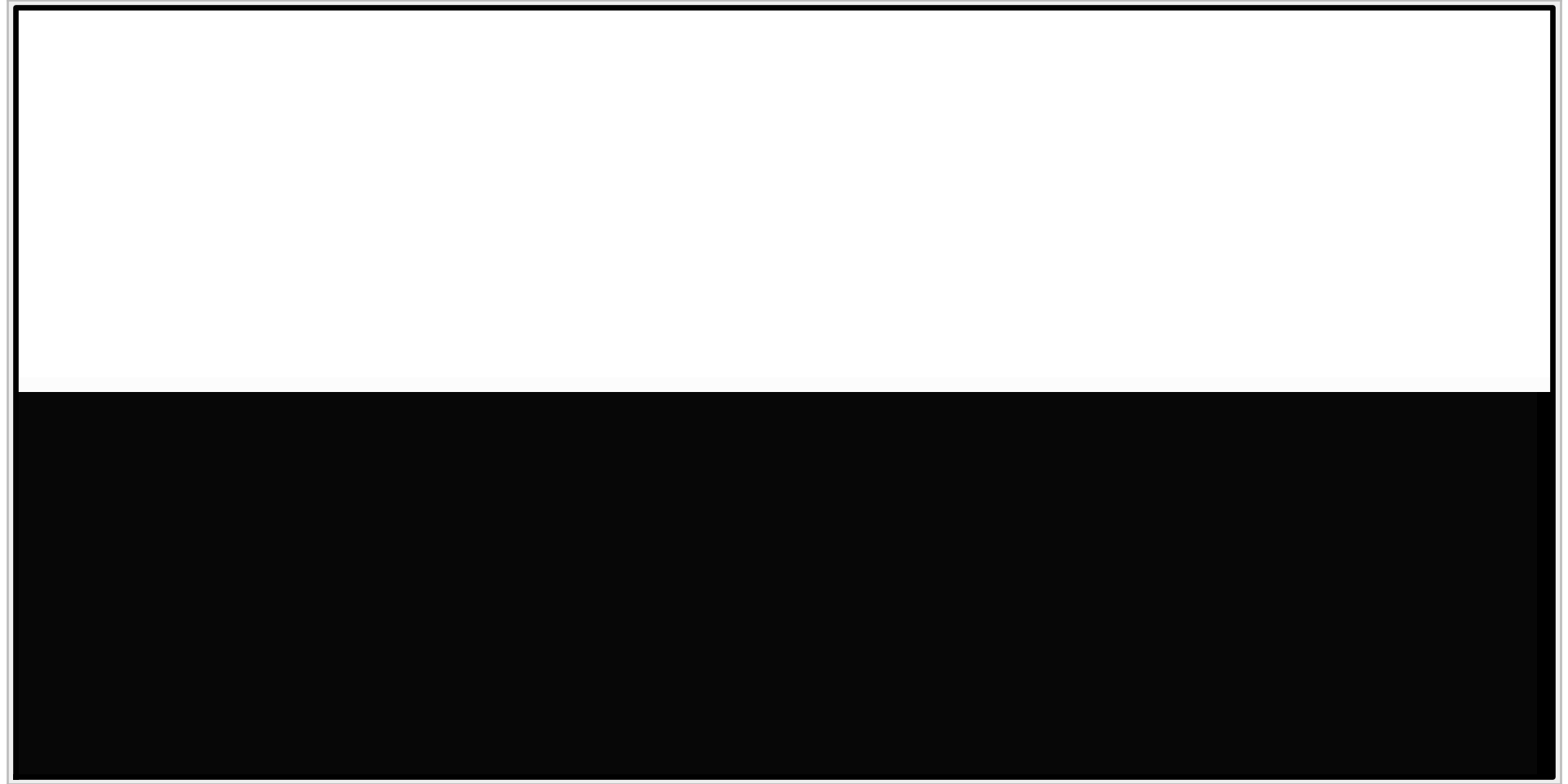}
    \includegraphics[width=0.32\textwidth]{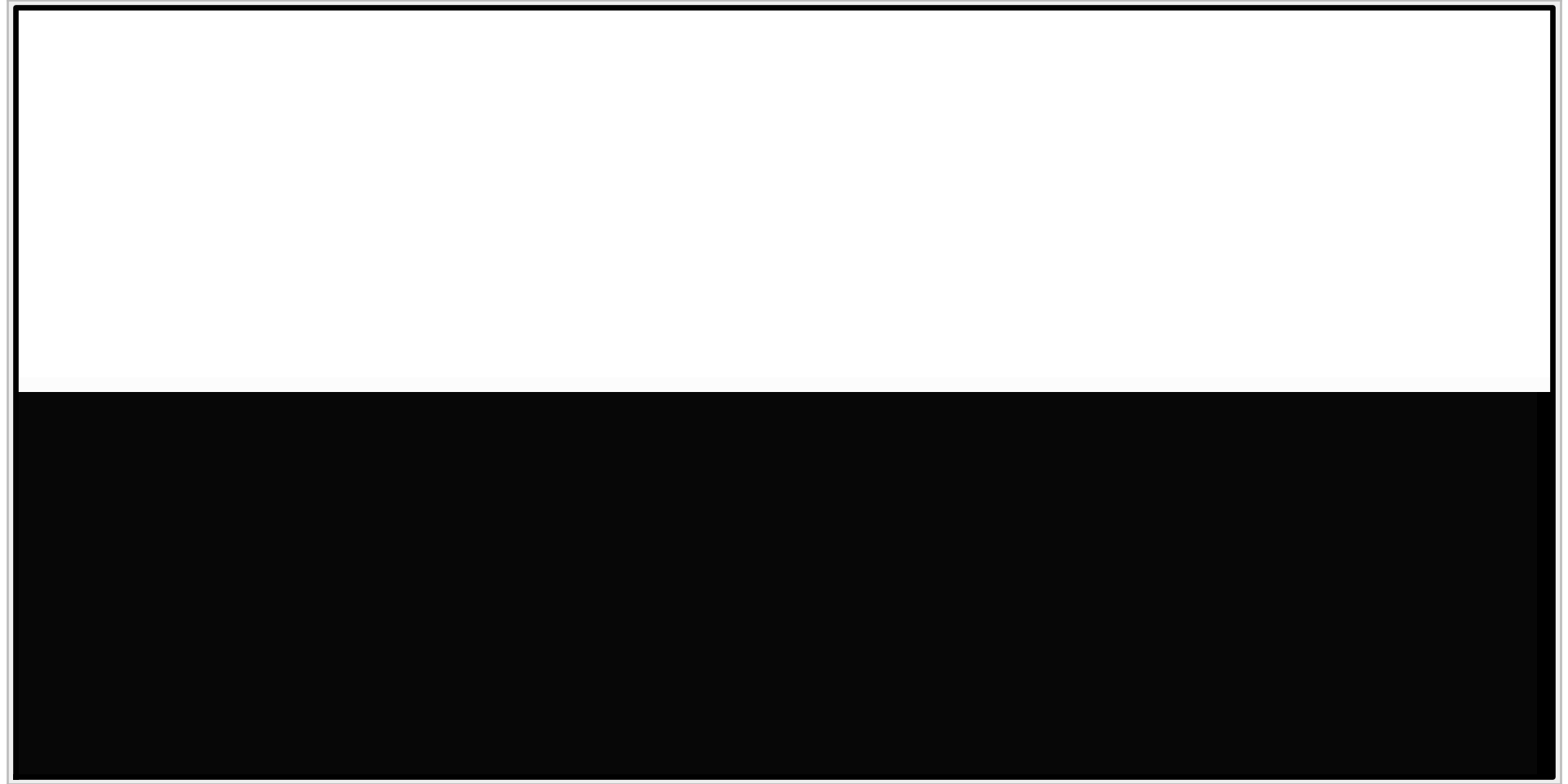}
    \includegraphics[width=0.32\textwidth]{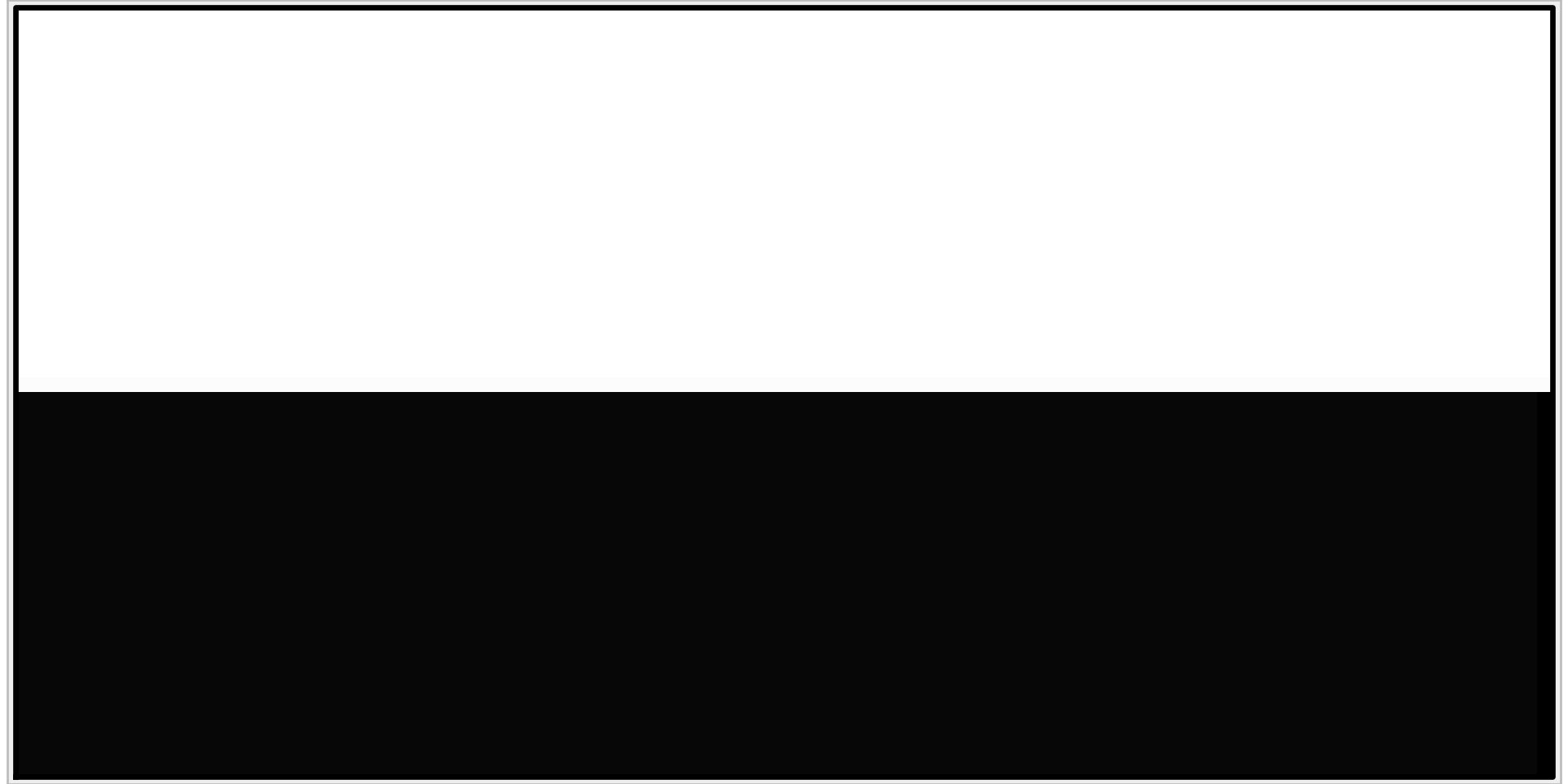}

    \includegraphics[width=0.32\textwidth]{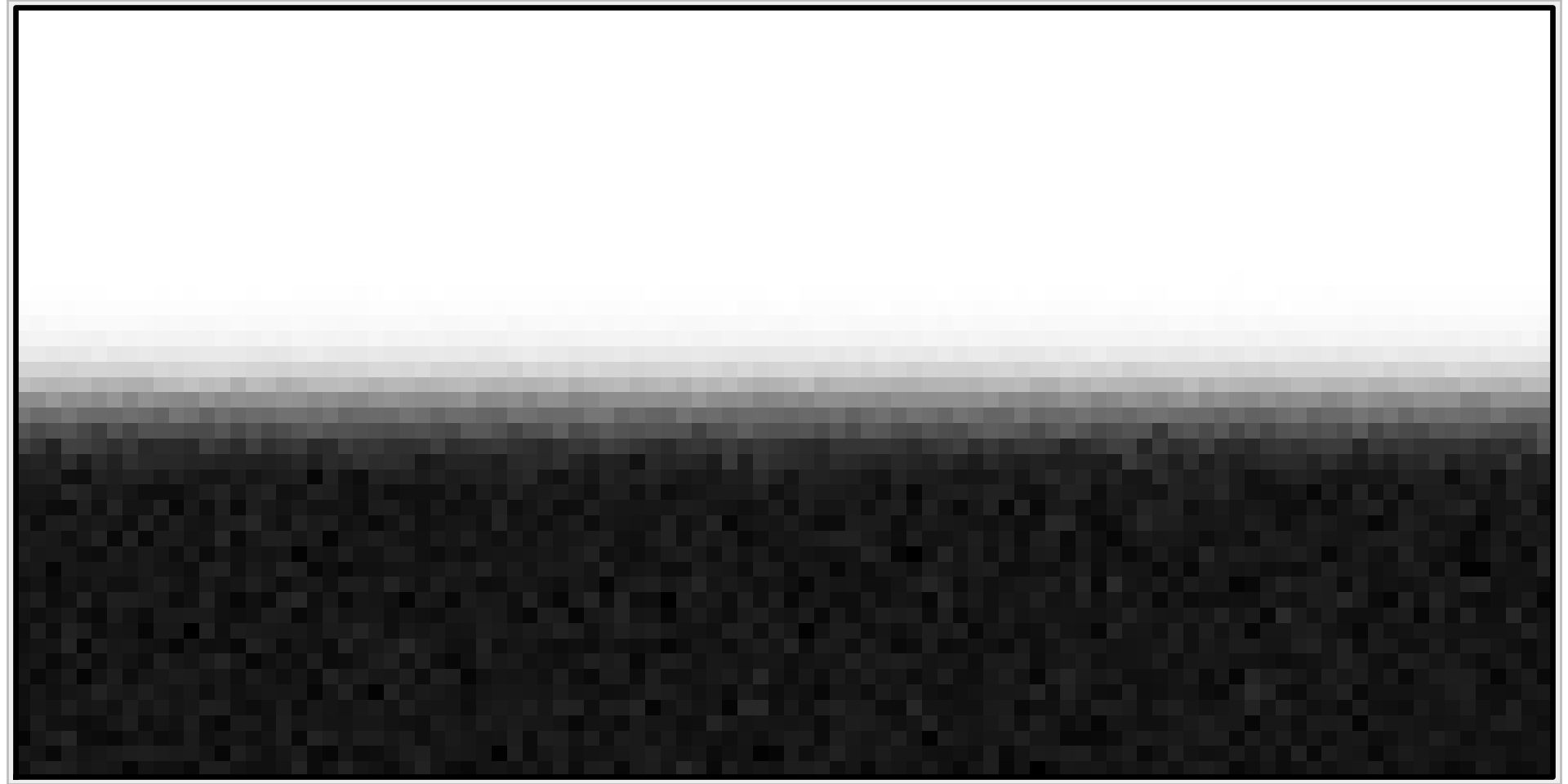}
    \includegraphics[width=0.32\textwidth]{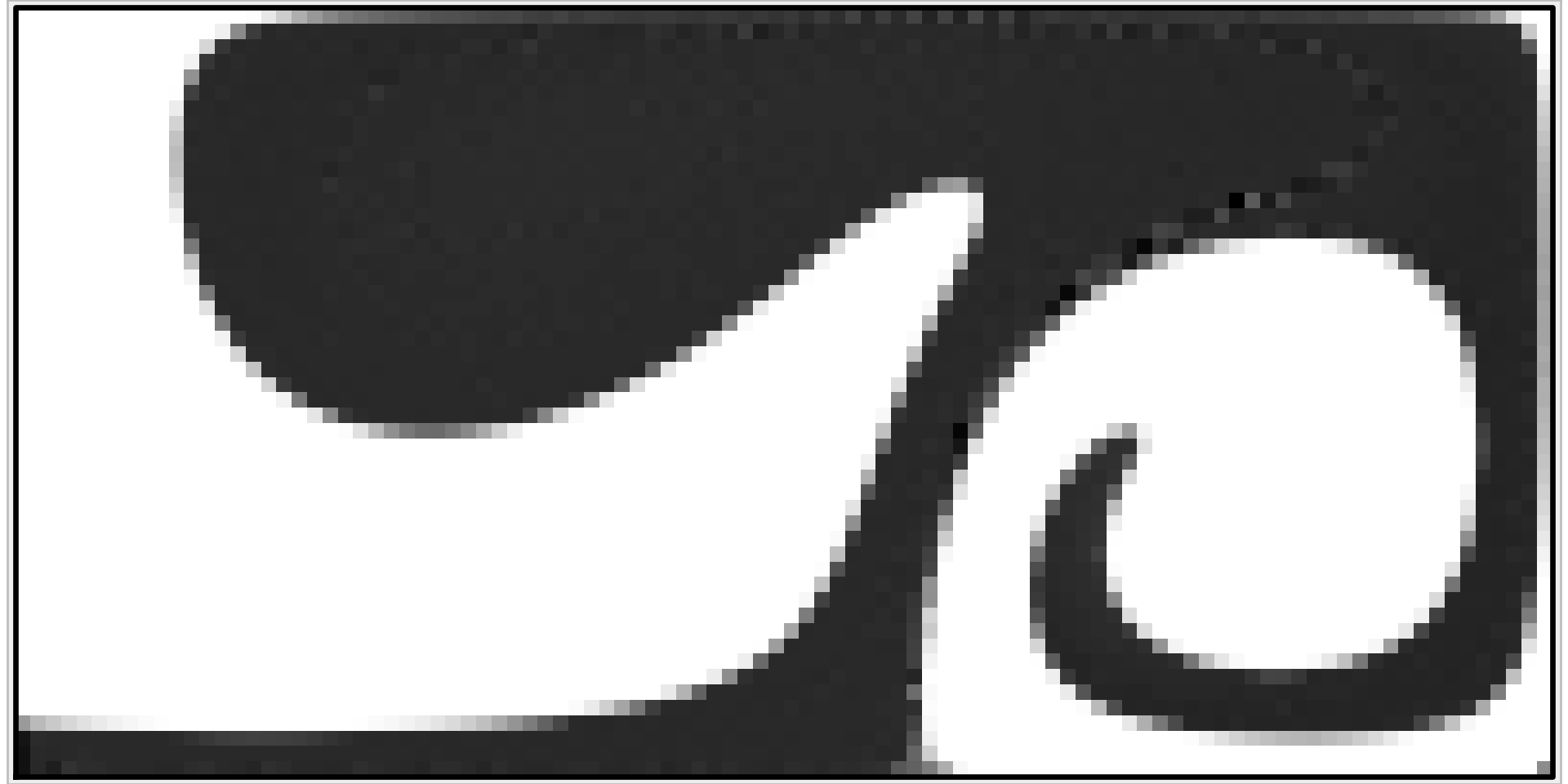}
    \includegraphics[width=0.32\textwidth]{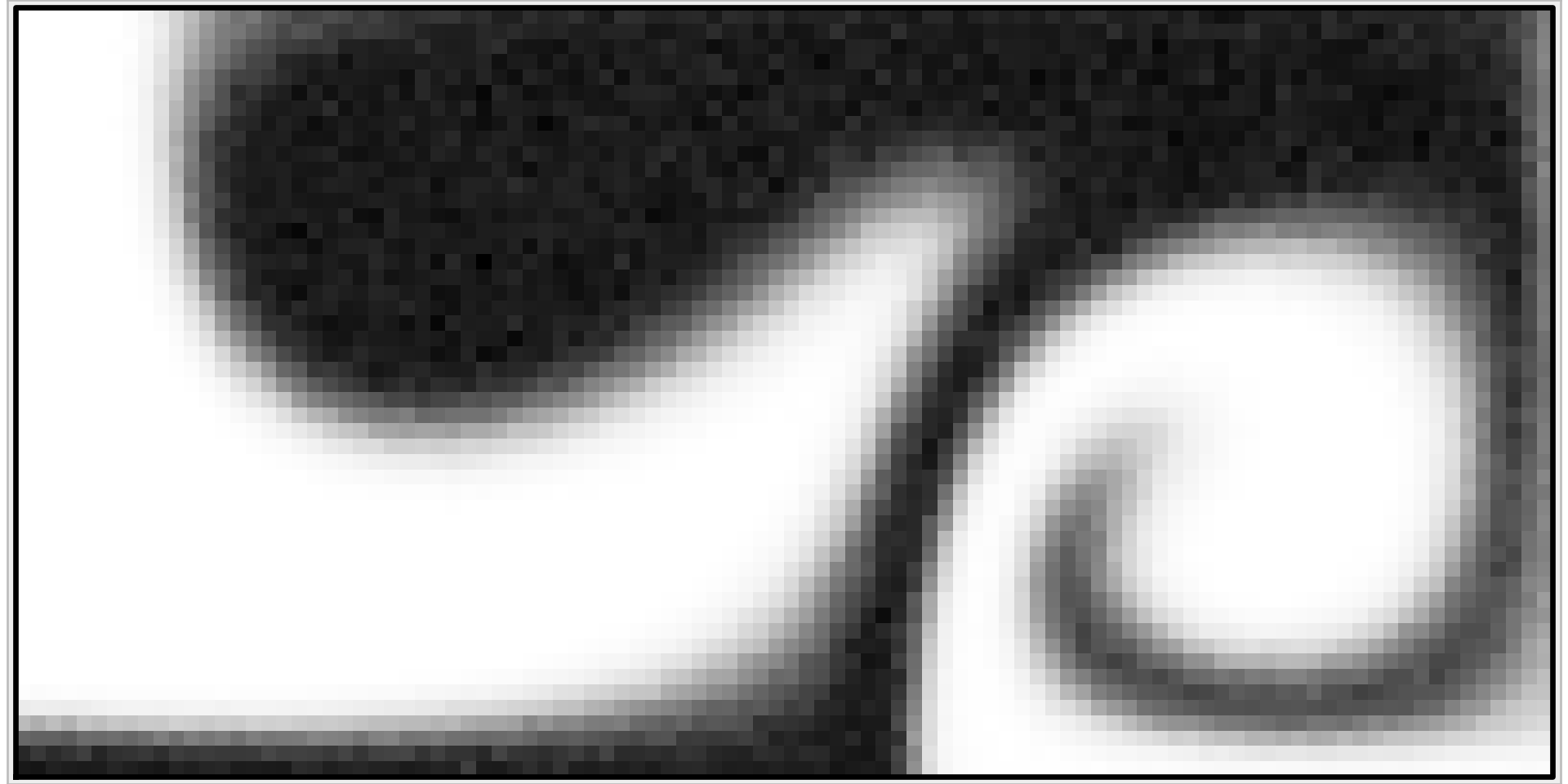}

    \includegraphics[width=0.32\textwidth]{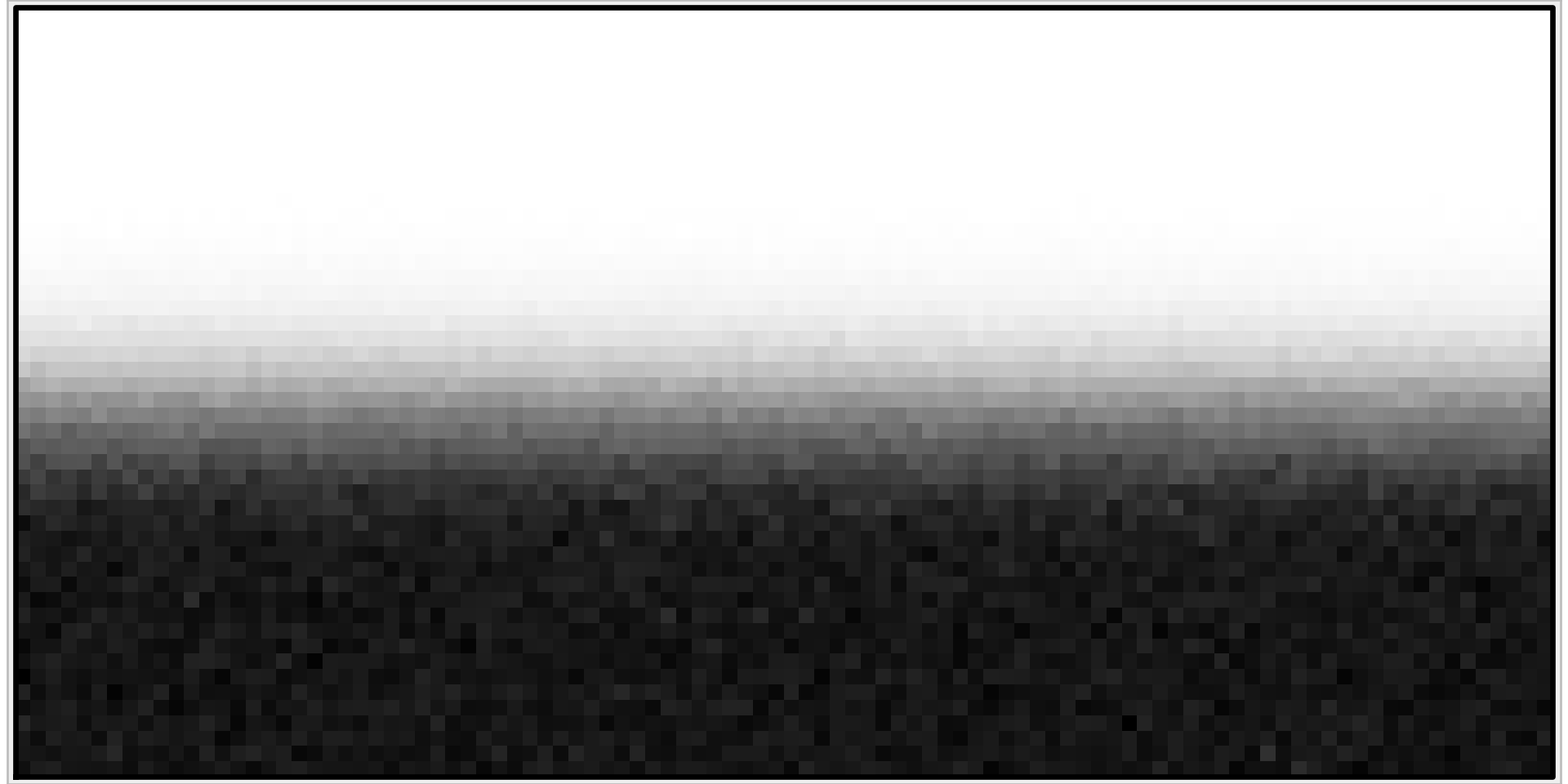}
    \includegraphics[width=0.32\textwidth]{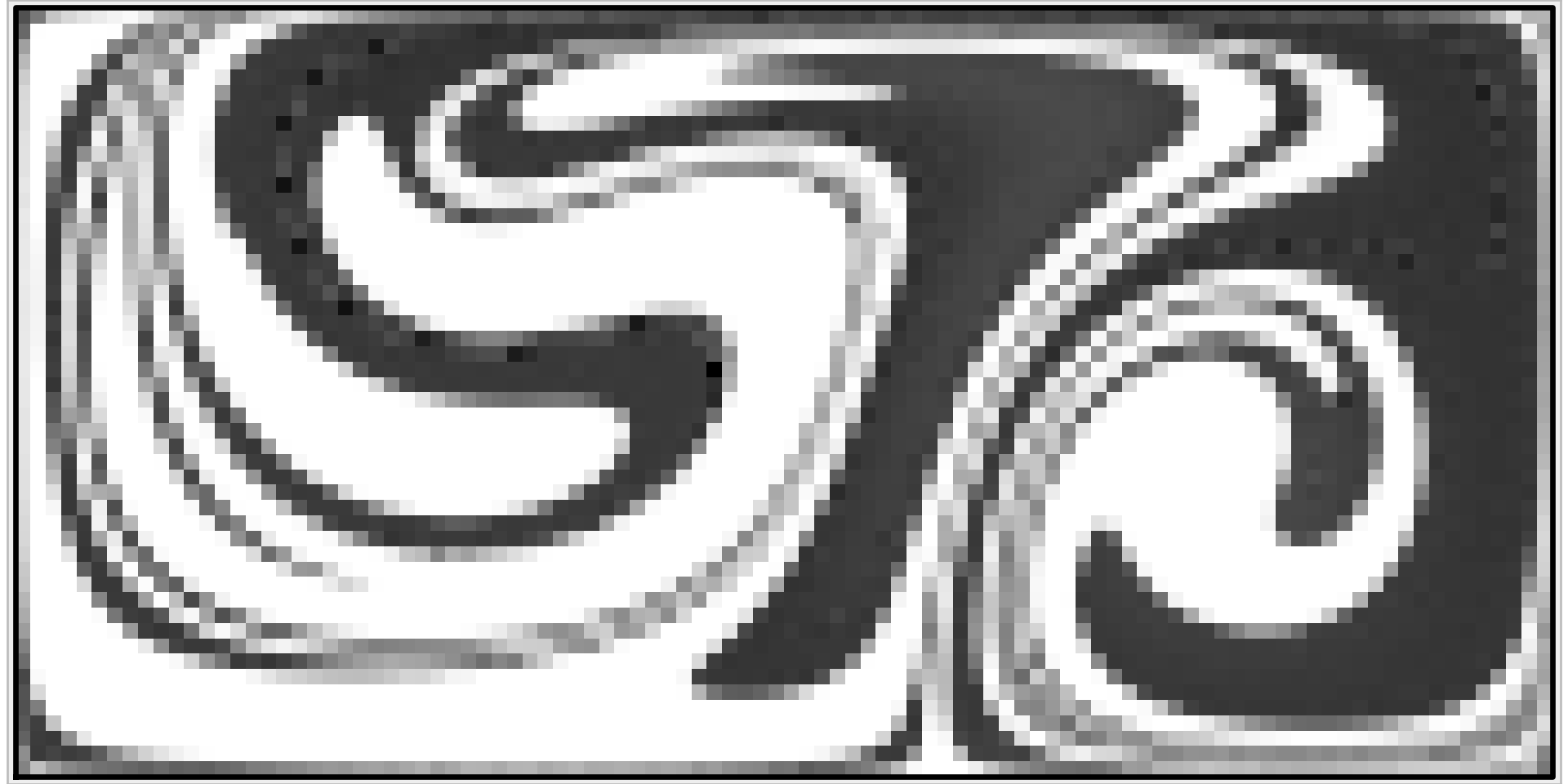}
    \includegraphics[width=0.32\textwidth]{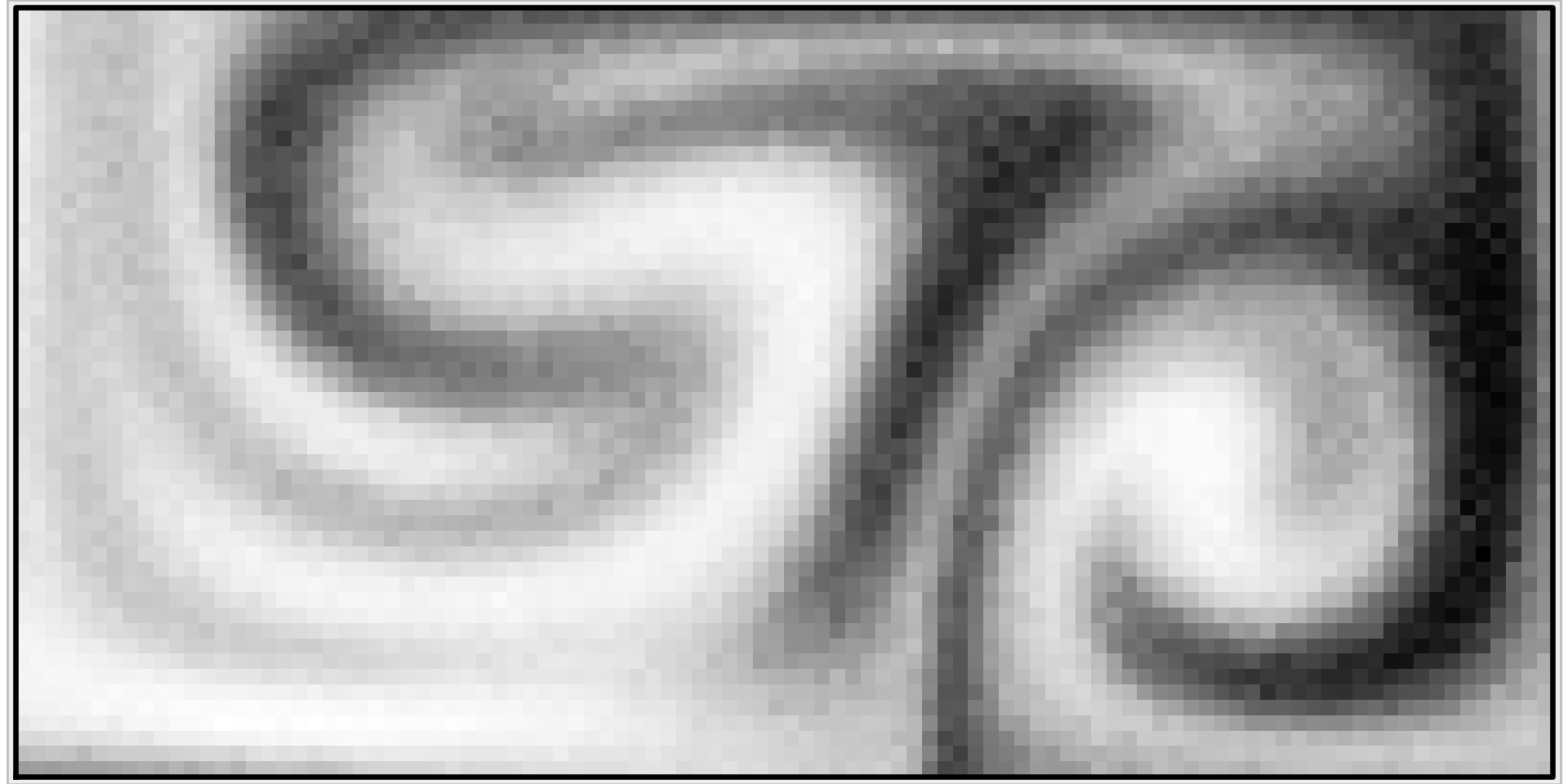}
    \end{center}
    \caption{Mixing of a passive tracer, initially located in the bottom half of the
    domain. In the left column, Fickian diffusion is applied, in the middle column,
    advection by a double gyre, and in the right column, both advection and
    diffusion.}
    \label{fig:turbulentmixing}
\end{figure*}

An illustration of this has been made in Fig.~\ref{fig:turbulentmixing}, where a
tracer initially located in the bottom half of a closed domain, is moved first
with diffusion only (Fig.~\ref{fig:turbulentmixing}, left column), then advected
by a double gyre (middle column), and finally both diffused and advected (right
column). While this is only intended as a schematic illustration, it shows how the
combined effect of advection by a gyre and Fickian diffusion leads to a faster
mixing than diffusion alone due to an increased interface between regions of
high and low concentration.

While ``turbulent mixing'' is in reality a combination of stirring by turbulence
and molecular diffusion, it will in almost all cases be impractical to model it
as such. In particular, eddies in the ocean exist at all scales, from the
largest ocean gyres with a scale of several thousand kilometers, to the smallest
turbulent eddies at the Kolmogorov scale \citep[p.~25]{davidson2015}, on the
order of 1 mm or less. Most numerical ocean models currently have a horizontal
resolution somewhere between 10s of meters and 10s of kilometers, and a vertical
resolution ranging from meters to 100s of meters. Any eddies smaller than the
resolution of the model cannot be resolved, and therefore their contribution to
the mixing must be parameterised, as the so-called eddy diffusivity. When
talking about diffusion in the context of oil trajectory modelling, and indeed
for the rest of this chapter, one typically refers to the eddy diffusivity.

\subsection{Origins of vertical mixing in the ocean}
\label{sec:origins_of_mixing}

% TODO add references to this section
There are many sources of turbulent kinetic energy (TKE) leading to vertical
mixing in the ocean. The most obvious (and spectacular) may be breaking waves,
which contribute to mixing in the upper part of the water column. Furthermore,
turbulent motion may be caused as currents flow across the sea bed, in narrow
straits, or due to shear flow between two fluid layers.  In cold or windy
conditions, evaporation or cooling at the surface will increase the density of
the water, and if the water at the surface becomes denser than the underlying
water, overturning will occur, leading to vertical mixing. In very cold
conditions, sea ice will form. During the freezing process, the salinity of the
ice is reduced through rejection of brine. This cold, high-salinity water will
have a density higher than the water below, again leading to overturning and
mixing of water masses.

In all cases, the stratification of the water column has a strong influence on
the vertical diffusivity. Near the surface, there will usually be a layer of
uniform density, called the surface mixed layer, or just the mixed layer. The
thickness, or depth, of the mixed layer will vary with geographical location,
time of year, and it will be influenced by local factors such as wind,
air temperature, rainfall and freshwater input from rivers.

While eddy diffusivity is typically high in the interior of the mixed layer, it
drops both close to the surface, and towards the bottom of the mixed layer.
Towards the surface, the mixing efficiency (which the eddy diffusivity
describes), is limited because the surface limits the size of turbulent eddies
\citep{craig1994}. This is a version of the so-called mixing length argument
(see, \emph{e.g.}, \citet[pp. 112--113]{davidson2015}), stating that the mixing depends
not only on the turbulent kinetic energy, but also on the size of eddies.

At the bottom of the mixed layer, the density increases, either due to an
increase in salinity, a decrease in temperature, or a combination of both. This
region of increasing density is called the pycnocline.  Stable stratification,
where a layer of light water overlays denser water, will tend to prevent mixing
across the density gradient, due to the energy required to lift the dense water
against gravity. For this reason, vertical diffusivity can be quite high
throughout the mixed layer, and then drop, sometimes by several orders of
magnitude, at the pycnocline. The interested reader is referred to
\citet{grawe2012} for further discussion of this particular case, in the context
of Lagrangian particle modelling.

As a side note, one might want to ask if the motion of the sea is actually
turbulent. If we make the approximation that the water in the mixed layer
behaves as an isolated slab of water, experiencing friction forces from the wind
at the top, and from the deep water at the pycnocline, then the Reynolds number
\citep[p.~23]{thorpe2005} for this flow is
\begin{align}
    \label{eq:Re_slab}
    \mathrm{Re} = \frac{\Delta u L}{\nu},
\end{align}
where $\Delta u$ is the difference between the speed at the top and the bottom
of the mixed layer, $L$ is the thickness of the mixed layer, and $\nu$ is the
kinematic viscosity of water, which is approximately
\SI[per-mode=symbol]{1.4e-6}{\meter\square\per\second} for seawater at
\SI{10}{\celsius}. Turbulent flow is commonly said to occur at Reynolds numbers
above approximately 4000. If we choose for example a mixed layer thickness of
$L=\SI{10}{\meter}$, we see that any velocity difference of more than about
$\SI[per-mode=symbol]{0.5e-3}{\meter\per\second}$ will give turbulent flow.

The above discussion of turbulent flow assumes that the density truly is
constant throughout the mixed layer. In regions where the density increases
slowly with depth, a relevant parameter to consider is the Richardson number,
\begin{align}
    \label{eq:Ri}
    \mathrm{Ri} =
        \frac{g}{\rho_w}
        \frac
            { \frac{\partial \rho_w}{\partial z} }
            { \left(\frac{\partial u}{\partial z}\right)^2 },
\end{align}
which is a dimensionless number related to the ratio between the stabilising
forces of stratification, and the destabilising forces of current shear. If
$\mathrm{Ri} \gg 1$, then the shear forces are not strong enough to break down
the density gradient and cause vertical mixing.

\subsection{Modelling ocean turbulence}
\label{sec:turbulence_modelling}

The vertical eddy diffusivity can be described through models of different
complexity, including simple parametrisations based on fitting simplified models
against experiments, and more complex models that try to solve dynamic equations
for transport and dissipation of turbulent kinetic energy.

The simplest model for vertical turbulent mixing would be to simulate a vertical
advection-diffusion process using a constant eddy diffusivity. However, in light
of the discussion in Section~\ref{sec:origins_of_mixing}, it should be clear
that this will in many cases be too simple. In particular, the diffusivity in
the mixed layer can often be several orders of magnitude higher than the
diffusivity at greater depths.

Hence, the second simplest approach might be to model the diffusivity profile as
a step function, with a high value in the mixed layer, and a lower value below
the pycnocline. However, note that care must be taken to avoid numerical
artifacts when using a step-function diffusivity. See
Sections~\ref{sec:stepfunction_example} and~\ref{sec:mixedlayer_example} for
examples and additional discussion of this topic.

Another approach is to use a simplified continuous model for the vertical
diffusivity. One such model, used in some oil spill modelling studies
\citep{skognes2004statmap, nordam2018oil}, is due to
\citet{ichiye1967}, who suggested the following relation for the vertical eddy
diffusivity as a function of depth ($z$ positive downwards):
\begin{align}
    \label{eq:ichiye}
    K(z) = 0.028 \frac{H_s}{T_p} \ue^{-2kz},
\end{align}
where $H_s$ is the significant wave height, $T_p$ is the peak wave period, and
$k$ is the wave number. This relation takes mixing due to waves into account,
but ignores the limiting effects of stratification, and does not feature reduced
diffusivity towards the surface.

A third option is to obtain eddy diffusivity from an ocean model. All or most
ocean models calculate eddy diffusivity, potentially taking into account waves,
density stratification, current shear, and more complex processes such as
Langmuir circulation \citep[pp. 251--255]{thorpe2005}. Several different
approaches at different degrees of complexity exist. So-called turbulence
closure schemes are a research field in themselves, and we will not go into
detail on the schemes themselves here. The interested reader is referred to,
\emph{e.g.}, \citet[p. 27]{davidson2015} and \citet[Chapter 5]{haidvogel1999}.

Relevant in the context of oil spill modelling is that many ocean models provide
eddy diffusivity as output on the same formats as the ocean current data. A
three-dimensional oil spill model will probably already be using currents from
an ocean model, and the advantage of using eddy diffusivity from the same model
is then that the fields are dynamically consistent. However, it is important to
be aware that the eddy diffusivity in an Eulerian ocean model also serves the
additional purpose of suppressing numerical instabilities that can occur in
advection-dominated problems.
%TODO find a reference for this claim
Hence, it is quite possible that the eddy diffusivity from an ocean model may be
somewhat higher than it should be, and therefore unsuited for direct use in a
Lagrangian transport model. Never the less, diffusivity from an ocean model
would be expected to take the important effects of stratification into account.

Finally, it is worth mentioning the existence of separate, stand-alone models
for vertical ocean turbulence. The most well-known of these is probably the
General Ocean Turbulence Model (GOTM) \citep{umlauf2005}\footnote{See also
\url{www.gotm.net}}. This is an open-source one-dimensional water column model,
that can be set up to model a range of different cases, with different forcings
as input, and using different turbulence closure schemes, such as
Mellor-Yamada \citep{mellor1982}, $k$-$\epsilon$ \citep{launder1983}, and
$k$-$\omega$ \citep{wilcox2008}. In an oil spill modelling
context, using a one-dimensional turbulence model is not as convenient as using
eddy diffusivity from an ocean model, but might be an option for a localised
area.

\subsection{Wave modelling}
\label{sec:waves}

As previously mentioned, breaking waves are a source of turbulent mixing in the
ocean. In the context of oil spill modelling, however, breaking waves are
perhaps even more important as the mechanism by which an oil slick at the
surface is broken up into droplets and submerged in the water column. We will
return to this point in Section~\ref{sec:entrainment}, but here we will mention
some approaches to obtain wave data for use in an oil spill model.

As with turbulence, there exists different approaches to obtaining wave data, at
different levels of complexity.  Advanced wave models that calculate the entire
wave spectrum exist, and may be run coupled to an ocean model (or
atmosphere-ocean model), such that the waves affect the calculation of the
current, and vice versa. An example of such a model is SWAN \citep{booij1997},
which may for example be coupled with the ROMS ocean model (see, \emph{e.g.},
\citet{warner2008}).

A simpler approach is to parameterise the wave state from the wind speed,
usually given at an altitude of 10 m above sea level. In the following example,
the significant wave height, and the peak wave period, $H_s$ and $T_p$, are
derived from the JONSWAP spectrum and associated empirical
relations~\citep{carter1982}. The sea state is assumed to be either
fetch-limited, or fully developed. Fetch-limited means that a steady state is
reached, where the sea does not reach a fully developed state because the fetch
(the distance over which the wind acts on the sea) is too short. This is
relevant close to the coast, in off-shore wind conditions. Fully developed, on
the other hand, refers to the steady state wave conditions that are found far
away from the coast. In both cases, the wave state is assumed \emph{not} to be
time-limited, \emph{i.e.}, the wind is assumed to have been constant for a
sufficiently long time to allow a steady wave state to develop.

Here, $H_s$ and $T_p$ are given by
\begin{subequations}
    \label{eq:wave_fetch_limited}
    \begin{align}
        H_s &= \frac{u_{10}^2}{g} H_c \sqrt{\frac{gL_f}{u_{10}^2}} \\
        T_p &= \frac{u_{10}}{g} T_{c} \sqrt[3]{\frac{g L_f}{u_{10}^2}}
    \end{align}
\end{subequations}
in the fetch-limited case, and
\begin{subequations}
    \begin{align}
    \label{eq:wave_fully_developed}
        H_s &= \frac{u_{10}^2}{g} H_{0} \\
        T_{p} &= \frac{u_{10}}{g} T_{0}
    \end{align}
\end{subequations}
in the fully developed case. Here, $H_{0} = 0.243$, $H_{c} = 0.0016$, $T_{0} = 8.134$, and $T_{c} = 0.286$ are dimensionless parameters, $g$ is the gravitational acceleration, $L_f$ is the fetch length, and $u_{10}$ is the wind speed at 10 m above sea level.

%%%%%%%%%%%%%%%%%%%%%%%%%%%%%%%%%%%%
%%%% Entrainment of surface oil %%%%
%%%%%%%%%%%%%%%%%%%%%%%%%%%%%%%%%%%%

\section{Entrainment of surface oil}
\label{sec:entrainment}

When a wave breaks on an oil slick, part of the oil in the breaking zone of the
wave will be entrained into the water column in the form of droplets. The amount
of entrained oil will increase with the height of wind-driven waves, and
therefore depends on the wind speed. To describe this in an oil spill model it
is necessary to formulate a model for the mass of oil entrained per unit of time
for a given surface slick in a given wave field. Some of the earliest
quantitative work on the entrainment of oil due to breaking waves is that of
\citet{delvigne1988}. In this classic paper, based on experiments in a
turbulence tank and two different meso-scale wave flumes (\SI{0.43}{\meter} and
\SI{4.3}{\meter} depth), they provided empirical relations for the three key
parameters in surface oil entrainment:
\begin{itemize}
    \item Droplet size distribution of the entrained oil,
    \item Entrainment rate,
    \item Intrusion depth. 
\end{itemize}
The relationships obtained by \citet{delvigne1988} were used for decades in oil
spill models, with new models only starting to take hold nearly thirty years
later. The empirical relationship for entrainment rate by Delvigne and Sweeney
was formulated in a convoluted way, where the entrainment rate depends on the
droplet size distribution, and the models lack theoretical support. These and
other aspects of the Delvigne and Sweeney models have been critisised by others
who have formulated alternative models in recent years
\citep{johansen2015,Li2016d}.

%%%%%%%%%%%%%%%%%%%%%%%%%%%%%%%%%%%%%%%%%%%
%%%%%%%% Droplet size distribution %%%%%%%%
%%%%%%%%%%%%%%%%%%%%%%%%%%%%%%%%%%%%%%%%%%%

\subsection{Droplet size distribution of entrained oil}
\label{ssec:dropletsizedistribution}

After entrainment of a surface slick, smaller droplets take longer to resurface
compared to larger droplets (see Section~\ref{sec:rise_speed}). For this reason,
the droplet size distribution of entrained oil is an important factor that
influences both the horizontal transport of the oil, and the lateral dispersion
from current shear. To describe dispersion, it is therefore necessary to use an accurate
droplet size model. Experimental evidence has shown that the size distribution
of droplets in a breaking wave event depends on oil viscosity
\citep{delvigne1988,reed2009}, oil-water interfacial tension
\citep{Zeinstra-Helfrich2016,li2017size}, oil film thickness
\citep{Zeinstra-Helfrich2016, Zeinstra-Helfrich2015c}, and energy in the
breaking wave \citep{delvigne1988}. Several published models exist which use
these and other parameters to estimate a droplet size distribution
\citep{delvigne1988,reed2009,Zhao2014a,johansen2015,li2017, nissanka2017}.

Two main droplet size model types can be distinguished. One type is an
equilibrium description, where the model consists of an expression for a
characteristic droplet size, such as the median, and a static droplet size
distribution function, such as a Rosin-Rammler, log-normal, or power-law
function, each with associated distribution parameters
\citep{delvigne1988,reed2009,johansen2015}. This formulation gives a static
distribution representative for some depth and after some time of wave impact.
The other type of formulation aims to calculate a dynamic droplet size
distribution through population balance models, which describe the
time-evolution of droplet breakup and coalescence in turbulent flow after wave
breaking \citep{Zhao2014a,nissanka2016calculation}. The equilibrium type model is
conceptually simpler and is easier to implement in an oil spill model, while the
population balance model may offer a more detailed description of the physical
process of droplet breakup. As of today, it is not clear which approach is best
for oil spill modelling.

In the following, the equilibrium type droplet size model of \citet{johansen2015}
will be described. This model is formulated from the observation that there are
two main regimes that determine the droplet size of oil in turbulent flow; a
viscosity-limited regime and an interfacial tension-limited regime. The first
regime is representative for weathered and emulsified surface oil, while the
second regime is representative for oil that has been treated with chemical
dispersants. Each regime is associated with the characteristic droplet size
through a non-dimensional number found from dimensional analysis. The
interfacial tension-limited regime is associated with the Weber number and the
viscosity-limited regime with the Reynolds number.

The Weber number is
\begin{align}
    \label{eq:We}
    \mathrm{We} = \frac{v^2 \rho_o h}{\sigma_{ow}},
\end{align}
where $\rho_o$ is the density of the oil, $\sigma_{ow}$ is the oil-water
interfacial tension, $h$ is the surface slick thickness, and $v = \sqrt{2 g H}$
a velocity scale related to the wave motion, with $H$ being the wave height.

The Reynolds number is given by
\begin{align}
    \label{eq:Re_pure}
    \mathrm{Re} = \frac{v \rho_o h}{\mu_{o}},
\end{align}
where $\mu_o$ is the dynamic viscosity of the oil.

Assuming that the characteristic droplet size can be found through a scaling
relationship involving these two numbers, in addition to three constants to be
determined from fitting to data, the ratio of characteristic droplet size $D$ to
slick thickness $h$ with the Reynolds and Weber numbers was found as
\begin{align}
    \label{eq:d50}
    \frac{D}{h} = A\We^{-a}\left[1+B^\prime\left(\frac{\We}{\ReNum}\right)^a\right]
\end{align}
The constants $A$, $B^\prime$ and $a$ appearing in this equation were fit
to experimental data in~\citet{johansen2015} as $A=2.251$, $B^\prime = 0.027$ and
$a=0.6$.

Equation~\eqref{eq:d50} provides a prediction for a characteristic droplet size
of the droplet size distribution. This can in principle be any characteristic
size and any distribution formulation; in \citet{johansen2015} these were taken
to be the median of the droplet size \emph{number} distribution, which was
described through a log-normal function. In oil spill modelling, the
\emph{volume} distribution for droplet sizes is needed, in order to account for
the mass of oil. From the log-normal distribution, one can obtain the volume
distribution from the number distribution by shifting the distribution as
described in \citet{johansen2015}. Specifically, the volume droplet size
distribution (for diameter $d$) is given as
\begin{align}
    \label{eq:sizedistribution} 
    v(d) &= \frac{1}{d\sqrt{2\pi}\sigma} \exp \left[-\frac{(\ln d - \mu)^2}{2\sigma^2} \right]
\end{align}
where we use a logarithmic standard deviation of
$\sigma=0.4\,\ln(10)$. The logarithmic mean $\mu$ is given by
$d^v_{50} = \ue^\mu$, and the relationship between the volume and number median
diameters is
\begin{align}
    \label{eq:median_diameters}
    \ln (d^v_{50}) = \ln (d^n_{50}) + 3 \sigma^2.
\end{align}

A similarly formulated model is the one of \citet{li2017}, which is a droplet
size distribution model intended to be valid for both surface entrainment by
breaking waves, and subsea blowouts through an orifice. In this model, the
maximum stable droplet size due to the Rayleigh-Taylor instability is used as a
length scale parameter, instead of the surface oil film thickness. Avoiding the
surface oil film thickness means that no separate model is needed to calculate
this value dynamically. At the same time, experimental evidence shows that
characteristic droplet size does scale with oil film thickness
\citep{Zeinstra-Helfrich2016}, making it an experimentally validated predictor,
although it should be noted that earlier work did not find a clear relationship
between the two variables \citep{delvigne1988}.

%%%%%%%%%%%%%%%%%%%%%%%%%%%%%%%%%
%%%%%%%% Entrainment rate %%%%%%%
%%%%%%%%%%%%%%%%%%%%%%%%%%%%%%%%%

\subsection{Entrainment rate of oil due to breaking waves}
\label{sec:entrainmentrate}

Historically, entrainment rate was explicitly or implicitly coupled to droplet
size distribution. One made the distinction between larger oil droplets that
almost immediately resurfaced after entrainment, and smaller droplets, that
became ``permanently entrained'' (see, \emph{e.g.}, \citet{reed1999} and
references therein). The net entrainment rate would then only
include the permanently entrained oil.

\citet[Section 4.4]{delvigne1988} found an expression for the entrainment rate
that explicitly included the droplet size: \begin{align}
\label{eq:entrainmentrate_DS} Q(d) = C \cdot D^{0.57}_{ba} S_{cov} F_{wc}
d^{0.7} \Delta d, \end{align} where $D_{ba}$ is the dissipated energy per
surface area [\SI{}{\joule\per\square\meter}], $S_{cov}$ is the sea surface area
fraction covered by oil, $F_{wc}$ is the sea surface area fraction hit by
breaking waves per second [\SI{}{\per\second}], $d$ is the droplet size
[\SI{}{\meter}] and $\Delta d$ is the width of the droplet size interval
(centered on $d$). The prefactor $C$ is an empirical constant that can include
the effects of oil state, such as viscosity, interfacial tension, and density;
in \citet{delvigne1988} only viscosity is included for different values of $C$.

It is not explicitly stated in the original work of \citet{delvigne1988} how
this equation should be applied to calculate the total entrainment rate; it has
however been interpreted in the literature \citep{li2017}. The equation gives
the entrainment rate over a droplet size interval, which means that an
integration over size intervals must be performed. This means that a lower and
upper limit of the droplet size must be decided upon.  It is likely that
different interpretations of Eq.~\eqref{eq:entrainmentrate_DS} exist, meaning
that models using this equation differently will provide somewhat different
results.

From a modelling point of view, a more elegant solution is to have an expression
of the entrainment rate that is completely independent of the droplet size
distribution, and heuristic concepts such as ``permanently dispersed oil''.
\citet{johansen1982} describes such an approach, modelling the vertical
transport (rise due to buoyancy, and vertical mixing due to eddy diffusivity)
with the advection-diffusion equation. He points out that in order to represent
a distribution of droplet sizes (and thus a distribution of rise speeds), one
must solve the advection-diffusion equation for each size class.

In such a model, the entrainment rate describes the amount
of oil that is submerged, and the droplet size distribution will describe how
that oil is distributed across size classes. The vertical transport model will
then determine the future development of those droplets, allowing large droplets
to surface rapidly, while small droplets remain submerged for longer periods.

Recent formulations of the entrainment rate adhere to this principle. Both
the two following examples describe the submersion of surface oil as a
first-order decay process
\begin{align}
    \label{eq:submersion_ODE}
    \frac{\ud Q_s}{\ud t} = -\alpha Q_s,
\end{align}
where $Q_s$ is the amount of oil at the surface, and $\alpha$ is the entrainment
rate.  \citet{johansen2015} describe a simple model where the submersion of
surface oil is related to the white-cap coverage fraction, $f_{wc}$ and the mean
wave period, $T_m$:
\begin{align}
    \label{eq:entrainmentrate_Johansen}
    \alpha = P f_{wc} / T_m.
\end{align}

In \citet{johansen2015}, Eq.~\eqref{eq:entrainmentrate_Johansen} was used as a
standalone model to describe the development of oil mass on the surface. It was
thus assumed that droplets larger than some limiting diameter would resurface
directly, and $P$ was then taken to be the fraction of droplets smaller than
this limiting diameter. For use in a modelling system where a vertical transport
model determines the fate of the droplets, we assume that oil is entrained at
the full rate, setting $P=1$. Then, the transport model will allow the larger
droplets to surface quickly.

\citet{li2017} developed an empirical relation parameterising the entrainment
rate, $Q$, in terms of the dimensionless Weber and Ohnesorge numbers:
\begin{align}
    \frac{Q}{F_{bw}} = a\mathrm{We}^b\mathrm{Oh}^c.
\end{align}
Here, $F_{bw}$ is the white-capping fraction per unit time [s$^{-1}$], the Weber
number is $\mathrm{We} = d_o \rho_w g H_s / \sigma_{ow}$, where $\rho_w$ is the
density of water, and the Ohnesorge number is $\mathrm{Oh} = \mu_o /
\sqrt{\rho_o \sigma_{ow} d_o}$. The length scale is the Rayleigh-Taylor
instability maximum droplet diameter, given by
\begin{align}
    \label{eq:rayleigh_taylor}
    d_o = 4 \left(\frac{\sigma_{ow}}{(\rho_w - \rho) g}\right)^{1/2}.
\end{align}
The values of the empirical parameters are $a=4.604\times10^{-10}$, $b=1.805$,
and $c=-1.023$~\citep{li2017}.

%%%%%%%%%%%%%%%%%%%%%%%%%%%%%%%%%%
%%%%%%%% Entrainment depth %%%%%%%
%%%%%%%%%%%%%%%%%%%%%%%%%%%%%%%%%%

\subsection{Entrainment depth of oil due to breaking waves}
\label{ssec:entrainmentdepth}

The linear parameterisation of intrusion depth and distribution of oil found by
    \citet{delvigne1988} is that after a wave breaking event, the oil is
    distributed evenly in the interval
\begin{align}
    \label{eq:intrusion}
    (1.5 - 0.35)H_w < z < (1.5 + 0.35)H_w,
\end{align}
where $H_w$ is the wave height. No more recent general model formulations for
the intrusion depth have been published in the oil spill literature. However, in
disagreement with this model, recent experiments in a wave tank gave intrusion
depth centers of less than half the wave height \citep{li2017size}. Other work
that may be relevant in this context includes studies of bubble entrainment by
breaking waves (see, \emph{e.g.}, \citet{leifer2006}) and observations of
vertical distributions of buoyant fish eggs (see, \emph{e.g.}
\citet{rohrs2014}).

%%%%%%%%%%%%%%%%%%%%%%%
%%%% Submerged oil %%%%
%%%%%%%%%%%%%%%%%%%%%%%

\section{Submerged oil}
\label{sec:submerged}

The vertical transport processes that affect submerged oil droplets are rise due
to buoyancy (or sinking in some cases, see, \emph{e.g.}, \citet{king2014}),
turbulent mixing, and vertical advection by currents. Of these three, advection
by vertical currents is probably the least important. Hence, we will not discuss
this further, other than to state that if current data with a vertical current
component is available, it can be used to advect the oil, just like the
horizontal components.

Vertical turbulent mixing has already been discussed in
Section~\ref{sec:turbulence_modelling}, and how to use the eddy diffusivity in
an oil spill model will be discussed in Sections~\ref{sec:eulerian}
and~\ref{sec:lagrangian}. Hence, the main content of this section will be the
calculation of rise speeds for oil droplets.

\subsection{Calculation of droplet rise speeds}
\label{sec:rise_speed}

It is commonly assumed that droplets, bubbles, sediment particles, etc.\ rise or
sink at their terminal velocity. The terminal velocity is derived by starting from the
observation that buoyancy exerts a constant force, $F_b$, on the submerged
particle:
\begin{align}
    \label{eq:buoyancy}
    F_b = \frac{4}{3}\pi r^3 g ( \rho_a - \rho_p ) = \frac{4}{3}\pi r^3 \rho_a g',
\end{align}
where $g' = g\frac{\rho_a - \rho_p}{\rho_a}$ is the reduced gravity, with
$\rho_a$ and $\rho_p$ the density of the ambient fluid and the moving particle,
respectively. While the buoyancy is constant, the drag force, $F_D$, increases
with the velocity, and has direction opposite to the velocity:
\begin{align}
    \label{eq:drag}
    F_D = -\frac{1}{2}\rho_a v^2 C_D A \cdot \frac{v}{|v|},
\end{align}
where $v$ is the velocity of the particle relative to the fluid, $A$ is the
cross-sectional area of the particle, and $C_D$ is a drag coefficient. By
setting the total force equal to 0, we get an equation that can be solved to
find the terminal speed, $v_b$:
\begin{align}
    \label{eq:risespeed}
    v_b = \sqrt{\frac{4}{3}\frac{d g'}{C_D}}.
\end{align}
The drag coefficient, $C_D$, is not constant, but rather a function of the
Reynolds number, which for a sphere is given by
\begin{align}
    \label{eq:Re}
    \ReNum = \frac{v d}{\nu_a} = \frac{\rho_a v d}{\mu_a}.
\end{align}
Here, $v$ is the speed of the sphere, $d$ is the diameter of the sphere, and
$\nu_a$ and $\mu_a$ are the kinematic and dynamic viscosities of the surrounding
fluid, and $\rho_a$ is its density.

At low Reynolds numbers, $\ReNum \ll 1$, the drag coefficient is given by
\begin{align} \label{eq:stokes_CD} C_D = \frac{24}{\ReNum}.  \end{align}
With this drag coefficient, the expression for the drag force becomes
\begin{align} \label{eq:stokes_force} F_D = -6 \pi r \mu_a v.  \end{align}
Solving for the terminal speed, $v_b$, one obtains \begin{align}
\label{eq:stokes} v_b = \frac{2}{9}\frac{\rho_p - \rho_a}{\mu_a}gr^2.
\end{align}
Equation~\eqref{eq:stokes_force} for the drag force is commonly known as Stokes'
law, after George Gabriel Stokes~\citep{stokes1856}, although
Eqs.~\eqref{eq:stokes_CD} and~\eqref{eq:stokes} are also sometimes referred to
as Stokes' law.

\begin{figure}
    \begin{center}
    \includegraphics[width=0.99\columnwidth]{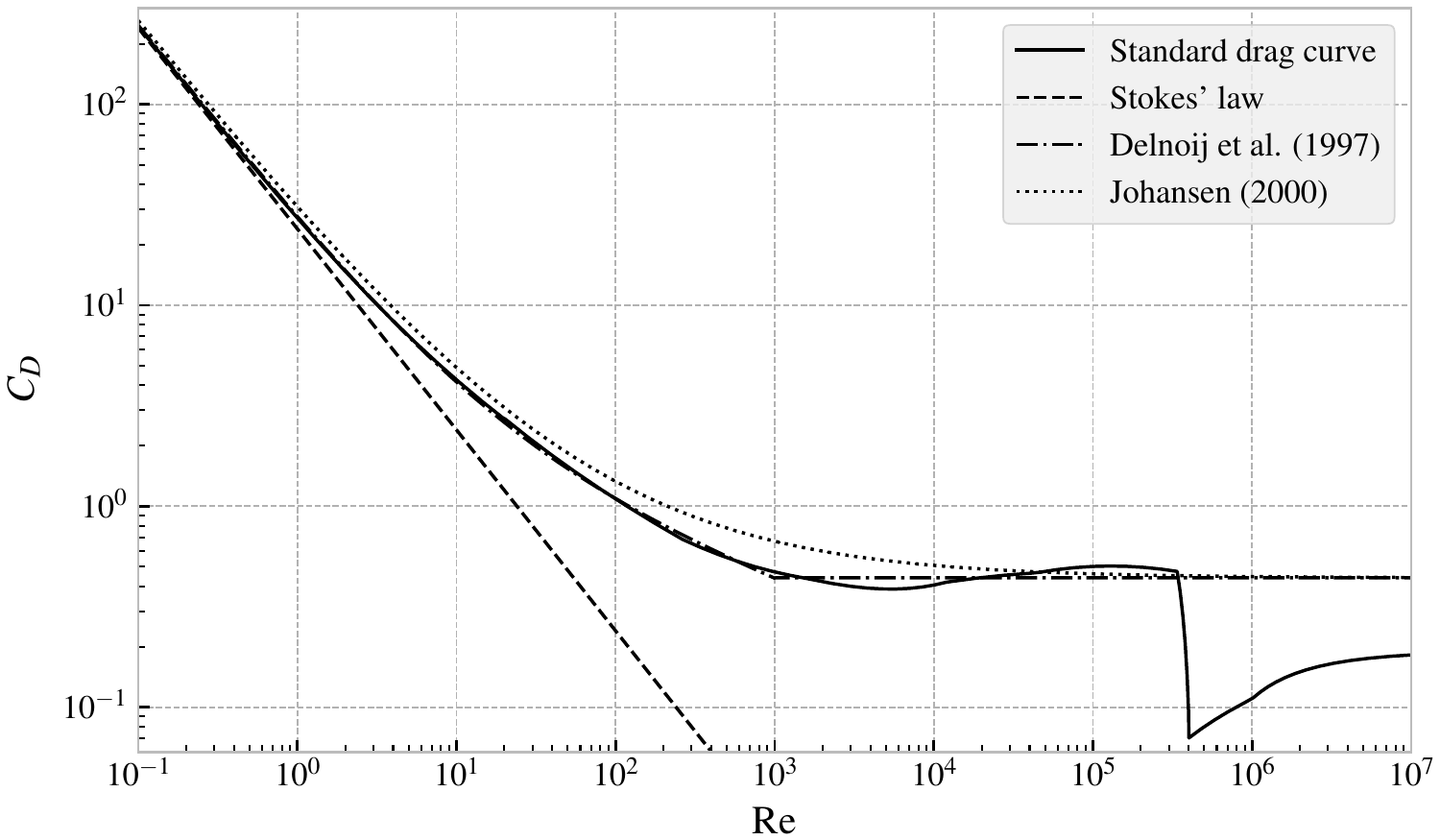}
    \end{center}
    \caption{Parameterisation of the Standard Drag Curve, due to \citet{clift1978}.
    Stokes' law, which is valid for $\ReNum \ll 1$, is shown as a dashed line.}
    \label{fig:standarddragcurve}
\end{figure}

In the derivation of Stokes' law, an assumption was made that the flow around
the spherical particle is laminar. At higher Reynolds numbers, the flow around
the sphere is no longer laminar, and Stokes' law no longer holds.  Various
empirical formulae exist for the case of high Reynolds number. \citet{clift1978}
combined several previously published results, and developed a piecewise
parameterisation of $C_D$ as a function of the Reynolds number, which they call
the Standard Drag Curve for the drag coefficient of a spherical
particle~\citep[pp. 110--112]{clift1978}. This parameterisation is shown in
Fig.~\ref{fig:standarddragcurve}, together with Stokes' law
(Eq.~\eqref{eq:stokes_CD}) and two other parameterisations.

As the highest range of Reynolds numbers in Fig.~\ref{fig:standarddragcurve} is
not relevant for oil droplets rising due to buoyancy, simpler expressions than
the standard drag curve have been suggested. \citet{delnoij1997} suggested a
parameterisation of $C_D$ given by
\begin{align}
    \label{eq:CD}
    C_D = \left\{
        \begin{array}{lll}
        \frac{24}{\ReNum}(1 + 0.15 \ReNum^{0.687}) & \mathrm{if} & \ReNum < 1000 \\
        0.44 & \mathrm{if} & \ReNum \geq 1000
        \end{array}
    \right. .
\end{align}
\citet{johansen2000} proposed a variant of the above scheme where the terminal
rise velocity is instead given by a harmonic transition between the high and low
Reynolds number cases:
\begin{align}
    \label{eq:johansen}
    v_b = \left( \frac{1}{v_1} + \frac{1}{v_2} \right)^{-1},
\end{align}
where $v_1$ and $v_2$ are calculated from Eq.~\eqref{eq:risespeed}, with a drag
coefficient of $C_D = 24/\ReNum$ in $v_1$, and $C_D = 0.44$ in $v_2$. The
parameterisations due to \citet{delnoij1997} and \citet{johansen2000} are also
shown in Fig.~\ref{fig:standarddragcurve}·

All of the above expressions assume spherical particles. In reality, an oil
droplet rising through water will deform to some degree, depending among other
things on the volume, density and oil-water interfacial tension.  Work on this
topic includes that of \citet{bozzano2001}, which takes droplet deformation into
account. They developed empirical relations for the drag coefficient and
deformation of droplets and bubbles in terms of the Reynolds, Morton and
E\"o{}tw\"o{}s numbers:
\begin{subequations}
    \begin{align}
        \label{eq:bozzano}
        C_D &= f \left(\frac{a}{R_0} \right)^2,
    \end{align}
where
    \begin{align}
        f &= \frac{48}{\ReNum}
        \left( \frac{1 + 12 \Mo^{1/3}}{1 + 36 \Mo^{1/3}} \right)
        + 0.9 \frac{\Eo^{3/2}}{1.4 (1 + 30 \Mo^{1/6})
        + \Eo^{3/2}}, \\
        \left( \frac{a}{R_0} \right)^2
        &= \frac{10(1 + 1.3 \Mo^{1/6}) + 3.1 \Eo} {10(1 + 1.3 \Mo^{1/6}) + \Eo},
    \end{align}
    and the Morton and E\"o{}tw\"o{}s numbers are given by
    \begin{align}
        \label{eq:MoEo}
        \Mo = \frac{g \mu_a^4 (\rho_a - \rho_p)}{\rho_a^2 \sigma^3},
        \;\;\;\; \Eo = \frac{g d_0^2 (\rho_a - \rho_p) }{\sigma}.
    \end{align}
\end{subequations}
Here, $\sigma$ is the interfacial tension between oil (or gas) and water, and
$d_0$ is the equivalent diameter of of the particle, \emph{i.e.}, the diameter of a
sphere with the same volume.

\subsection{Role of dispersants}
\label{sec:dispersants}

Oil dispersants are specifically designed surfactant chemicals intended to
reduce the oil-water interfacial tension. Dispersants can be used as a
countermeasure during oil spill response, with the objective of enhancing the
dispersion of the spill by facilitating the creation of small oil droplets. This
can be done subsea during a blowout, where the dispersants are injected directly
into the oil stream, facilitating breakup of the oil into smaller droplets in
the turbulent jet, or it can be done at the surface, where the treated oil will
be broken up into small droplets when hit by breaking waves or other mechanical
energy.

Looking at Eqs.~\eqref{eq:We} and~\eqref{eq:d50}, we see that reduced
interfacial tension gives a larger Weber number, which in turn gives a smaller
characteristic droplet size in natural dispersion, when everything else is kept
constant. While droplet breakup in turbulent jets is outside the scope of this
chapter, we can briefly mention that also in this case the droplet size be may
related to the Weber number, and reduced interfacial tension with everything
else kept constant will lead to smaller droplets~\citep{brandvik2013,
johansen2013}.

Small droplets produced by dispersant application will have a slower rice
velocity, as discussed earlier in this chapter. As also mentioned, this will
lead to the oil being spread over a larger volume of water, due to current shear
and horizontal transport and mixing. Smaller droplets give rise to faster
dissolution and faster biodegradation of the oil, potentially decreasing the
overall lifetime of contamination after the spill.  However, one should also be
aware that applying dispersants does not remove the oil, even if it is less
visible. Dispersant application can reduce the potential impact to sea birds,
mammals, and shoreline habitats, but at the cost of potentially increasing the
impact on marine life in the water column, as well as benthic species.

%%%%%%%%%%%%%%%%%%%%%%%%%%%%%%%%%%%%%%%%%%%
%%%% Eulerian model of vertical mixing %%%%
%%%%%%%%%%%%%%%%%%%%%%%%%%%%%%%%%%%%%%%%%%%

\section{Eulerian model of vertical mixing}
\label{sec:eulerian}

Calculating in the Eulerian picture means to consider concentrations at a set of
points (or in a set of cells), and looking at how the concentrations in those
points change with time. Solving a partial differential equation (PDE), such as
the advection-diffusion equation, for a discrete grid of points, is an example
of an Eulerian calculation. For different reasons, which will be described in
more detail in Section~\ref{sec:lagrangian}, it is not very common to solve oil
spill problems in the Eulerian picture.  Nevertheless, some background
information on the Eulerian picture is very useful, as this forms the starting
point for the Lagrangian, particle based approach of oil spill modelling.

\subsection{Advection-diffusion equation}
\label{sec:ad}

The change in concentration, along the vertical dimension, of oil droplets that
rise due to buoyancy and are mixed due to ocean turbulence, is commonly modelled
as an advection-diffusion problem.  Assuming the droplets to rise with a
constant, terminal velocity, $v_b$, and that the spatially dependent diffusivity
can be expressed as a function of depth and time, $K(z, t)$, the concentration
of droplets as a function of space and time, $C(z, t)$, is described by
\begin{align}
    \label{eq:ad-equation}
    \frac{\partial C}{\partial t}
    = \frac{\partial}{\partial z}\left( K \frac{\partial C}{\partial z} \right)
    - v_b \frac{\partial C}{\partial z}.
\end{align}
If we have $v_b = 0$ and $K(z, t) > 0$, then this equation is simply the
diffusion equation (also known as the heat equation). For simple geometries and
initial conditions, analytical solutions are known for many cases, in particular
if $K(z, t)$ is a constant (see \emph{e.g.}, \citet{carslaw1959})
With $v_b \neq 0$ and $K(z, t) = 0$, Eq.~\eqref{eq:ad-equation} becomes the
advection equation, which describes the transport of a concentration profile,
without diffusion. For the special case of constant $v_b$, the advection
equation in one dimension describes a concentration profile that moves at a
speed of $v_b$, without changing its shape.

For practical applications, it is usually not possible to find analytical
solutions to Eq.~\eqref{eq:ad-equation}. In such cases, a range of numerical
solution techniques exist. For details, the interested reader is referred to the
wide range of literature on the topic of numerical solutions of PDEs. See,
\emph{e.g.}, \citet{hundsdorfer2003, versteeg2007, pletcher2013}.

\subsection{Boundary conditions}
\label{sec:eulerBC}

In oil spill modelling, it is essential to distinguish between surface oil and
submerged oil. Surface oil is not only distinguished by being located at zero
depth, but also by the fact that surface oil is not subject to vertical mixing
due to turbulence in the water column. The idea behind this is that submerged
oil are found in the form of droplets, which are surrounded by water and thus
subject to turbulent motion. Surface oil, on the other hand, is present in the
form of continuous patches of different sizes. In order to submerge oil from a
patch or slick at the surface, some high-energy event, such as a breaking wave,
is required to break the surface tension of the oil. The opposite process,
\emph{i.e.}, surfacing, is usually calculated from the buoyant rise speed of the
oil droplets. Oil that reaches the surface due to buoyancy may leave the water
column and merge with the surface slick. This makes oil behave differently than,
\emph{e.g.}, buoyant fish eggs, as these do not ``get stuck'' at the surface in
the same way \citep{sundby2015principles}.

One way to model this behaviour of oil is to assume that the concentration in
the water column is described by the advection-diffusion equation, with a
partially absorbing boundary at the surface. In particular, the boundary at the
surface should enforce zero diffusive flux, while allowing the advective flux
due to buoyancy to leave the water column through the boundary at the surface.
The oil that has left the water column in this manner is then counted as part of
the surface oil. The physical rationale for this choice of boundary conditions
is that higher buoyancy (due to either larger droplets or less dense oil)
\emph{does} lead to faster surfacing, while higher diffusivity \emph{does not}
lead to faster surfacing.

The advective and diffusive fluxes are given by:
\begin{subequations}
    \label{eq:fluxes}
    \begin{align}
        \label{eq:jA}
        j_A(z, t) &= w C(z, t), \\
        \label{eq:jD}
        j_D(z, t) &= - K(z, t) \frac{\p C(z, t)}{\p z},
    \end{align}
\end{subequations}
where Eq.~\eqref{eq:jD} is commonly known as Fick's law (see, \emph{e.g.}, \citet[p.
4]{csanady1973}). Hence, a no-diffusive-flux boundary condition at $z=0$ can be
enforced by requiring
\begin{align}
    \label{eq:ndfBC}
    \left. \frac{\p C(z,t)}{\p z} \right|_{z = 0} = 0.
\end{align}

Another option for modelling the surfacing of oil is to consider the surfacing
process as a loss term in the PDE (also known as a sink), where oil which is
sufficiently close to the surface is simply removed, at a rate which would
typically be calculated from the rise speed of the oil droplets. See,
\emph{e.g.}, \citet{tkalich2002} for an example of this approach. Note that the
same rate of surfacing can be modelled in both approaches.

When considering a finite water depth, the boundary at the bottom should also be
reflecting for the diffusion step. As long as the oil considered is positively
buoyant, the advective flux through the bottom will necessarily remain zero. In
advanced oil spill models, interaction with seabed sediments of different types
through a turbid bottom layer may be included, where adhesion of oil to sediments
is explicitly modelled. One may also wish to account for the possibility of
sinking droplets of oils that are denser than water settling onto the sediment.
%TODO insert references for this claim

\subsection{Source term for entrainment of oil}
\label{sec:source}

In the scheme described above, the concentration of submerged oil in the water
column is described by the advection-diffusion equation
(Eq.~\eqref{eq:ad-equation}). In this case, we may model the entrainment of oil
by adding a reaction term to Eq.~\eqref{eq:ad-equation}, which adds oil at
certain depths. For example, if oil is entrained at rate $Q(t)$ (units mass per
time), and distributed evenly across a depth interval ranging from $H_{min}$ to
$H_{max}$, we have
\begin{subequations}
    \begin{align}
        \label{eq:ad-equation_with_source}
        \frac{\partial C}{\partial t}
        = \frac{\partial}{\partial z}\left( K(z) \frac{\partial C}{\partial z} \right)
        - v_b \frac{\partial C}{\partial z} + R(z, t),
    \end{align}
    where
    \begin{align}
        \label{eq:source}
        R(z, t) = \left\{
            \begin{array}{ccc}
                Q/L & \mathrm{if} & H_{min} < z < H_{max} \\
                0 & \mathrm{otherwise}
            \end{array}
        \right. ,
    \end{align}
\end{subequations}
where $L = H_{max} - H_{min}$.

\subsection{Modelling a droplet size distribution}
\label{sec:eulerian_distribution}

An important point in oil spill modelling is the concept of a droplet size
distribution, as discussed in Section~\ref{ssec:dropletsizedistribution}. As oil
is submerged due to breaking waves, a range of droplet sizes are produced, and
these will have a different fate in the water column. Not only does the droplet
size strongly influence the rise speed (see Section~\ref{sec:rise_speed}), but
also dissolution and biodegradation are affected by the droplet size, due to the
change in surface area relative to volume.

To capture the effect of droplet size on rise velocity in an Eulerian model, one
needs to separate the submerged oil into discrete droplet size classes, and
solve one advection-diffusion equation for each size class. No exchange between
droplet size classes is necessary for the submerged oil, but the source term
(Eq.~\eqref{eq:source}) must be modified such that the correct proportion of the
submerged oil is inserted into each size class \citep{kristiansen2020}.

\subsection{The well-mixed condition}
\label{sec:wmc}

The well-mixed condition (WMC), described by \citet{thomson1987}, states that a
passive (\emph{i.e.}, neutrally buoyant) tracer that is initially well mixed,
must remain well mixed while undergoing diffusion.  This holds regardless of the
shape of the diffusivity profile, and provided of course that the tracer cannot
escape through domain boundaries or similar. The well-mixed condition follows
directly from the diffusion equation for a concentration, $C(z, t)$:
\begin{align}
    \label{eq:diffusion}
    \frac{\partial C}{\partial t}
    = \frac{\partial}{\partial z}\left( K(z) \frac{\partial C}{\partial z} \right).
\end{align}
If $\partial_z C(z, t) = 0$ everywhere (including at the boundaries), then the
right-hand side of Eq.~\eqref{eq:diffusion} is 0, and hence there will be no
change in concentration as time passes.

In Eulerian modelling of diffusion, it is fairly straightforward to ensure that
the WMC is satisfied. In Lagrangian modelling, on the other hand, this is not
always trivial. However, as stated by \citet{thomson1987}, the WMC is a
necessary (though not sufficient) condition for a Lagrangian scheme to be
consistent with the diffusion equation. We will return to this point in
Section~\ref{sec:lagrangian}.

%%%%%%%%%%%%%%%%%%%%%%%%%%%%%%%%%%%%%%%%%%
%%%%%%%%%% Lagrangian modelling %%%%%%%%%%
%%%%%%%%%%%%%%%%%%%%%%%%%%%%%%%%%%%%%%%%%%

\section{Lagrangian modelling of vertical mixing}
\label{sec:lagrangian}

In oil spill modelling, the most common approach to simulating the transport and
mixing of oil at sea is to represent the oil as numerical particles, also called
Lagrangian elements (or sometimes ``spillets''). These numerical particles move
with the current, rise or sink according to their buoyancy, and move randomly to
account for turbulent mixing. When a large number of Lagrangian elements is
simulated, their distribution may be used to approximate the concentration field
of a substance, such as oil.

In this section, we describe some of the theory behind the use of particles to
model an advection-diffusion problem, and some conditions that must be satisfied
in order for this approach to be equivalent to the Eulerian approach described
above. We also give numerical schemes for the transport and mixing, the boundary
conditions, and the entrainment.

The link between the diffusion equation, and the distribution of a collection of
randomly moving particles, has been known for a long time. More than 100 years ago,
\citet{einstein1905} showed that the random motion of Brownian particles
(\emph{e.g.}, tiny pollen grains suspended in water) caused them to spread out
in accordance with the diffusion equation on long time scales. A few years later,
\citet{langevin1908} presented a differential equation for the movement of
Brownian particles, based on Newton's second law with a stochastic force term.

Since then, the mathematical field of stochastic differential equations has been
developed further, and put these results on a more solid theoretical foundation.
In the following, we shall only use a few elements of the theory of stochastic
differential equations, but references to further reading will be given where
relevant.
%TODO add references to Itô and Kolmogorov (and others?)

\subsection{Modelling vertical diffusion as a random walk}
\label{sec:randomwalk}

Diffusion in a Lagrangian model is described by a random walk, \emph{i.e.}, a
random displacement of particles at each timestep. More formally, a random walk
is an example of a Stochastic Differential Equation (SDE), which is a
differential equation that includes one or more random terms. A general
one-dimensional SDE with one noise term is written
\begin{align}
    \label{eq:sde_general}
    \ud z = a(z, t) \, \ud t + b(z, t) \, \ud W_t,
\end{align}
where $a(z, t)$ is called the drift term, $b(z, t)$ is called the diffusion term
or noise term, and $\ud W_t$ are the random increments of a standard Wiener process,
$W(t)$ \citep[p. 40]{kloeden1992}. 

To solve this equation numerically, we first introduce a discrete time,
\begin{align}
    \label{eq:tn}
    t_n = t_0 + n\Delta t,
\end{align}
and then we seek a scheme to calculate the next position, $z_{n+1}$, given the
position, $z_n$, at time $t_n$. Numerous numerical schemes for SDEs exist, the
simplest of which is the Euler-Maruyama scheme \citep[p.
305]{maruyama1955,kloeden1992}. In this scheme, the iterative procedure for
integrating Eq.~\eqref{eq:sde_general} reads
\begin{align}
    \label{eq:euler_scheme}
    z_{n+1} = z_n + a(z_n, t_n) \, \Delta t + b(z_n, t_n) \, \Delta W_n,
\end{align}
where $z_n$ is the position at time $t_n$, and $\Delta W_n$ is a random number
drawn from a Gaussian distribution with zero mean, $\langle \Delta W \rangle = 0$,
and variance $\langle \Delta W^2 \rangle = \Delta t$.

For our purposes, it can be shown that if one solves the following
SDE for a large number of particles,
\begin{align}
    \label{eq:sde}
    \ud z = (w + K'(z))\, \ud t + \sqrt{2 K(z)}\, \ud W(t),
\end{align}
then their distribution will develop
according to the advection-diffusion equation (Eq.~\eqref{eq:ad-equation}), with
advection $w$, and diffusivity $K(z)$. Additionally, in Eq.~\eqref{eq:sde}, 
\begin{align}
    \label{eq:prime}
    K'(z) = \left.\frac{\p K}{\p z}\right|_z.
\end{align}
See \ref{app:equivalence} for details on how to derive
Eq.~\eqref{eq:sde} from Eq.~\eqref{eq:ad-equation}.

If we let the advection term be equal to the rise speed due to buoyancy, $w = v_b$,
and discretise Eq.~\eqref{eq:sde} with the Euler-Maruyama scheme, we obtain
\begin{align}
    \label{eq:euler}
    z_{n+1} = z_n + \big( v_b + K'(z_n) \big)  \, \Delta t + \sqrt{2K(z_n)} \, \Delta W_n.
\end{align}
This equation is a discrete formulation of the transport equation for
numerical particles. A similar expression may be used for the horizontal
directions. Some variant of this equation is commonly seen in papers on
numerical oil spill modelling. Note, though, that it is also quite common to see
this equation \emph{without} the term $K'(z_n) \, \Delta t$, in which case it is
\emph{not} consistent with the advection-diffusion equation (except in the
special case where $K$ is a constant). See Section~\ref{sec:naive} for further
details.

Just as for Ordinary Differential Equations (ODEs), a range of different
numerical schemes exist for solving SDEs such as Eq.~\eqref{eq:sde}. For a
review of several different schemes in the context of marine particle transport,
see \emph{e.g.}, \citet{grawe2011, grawe2012}. The interested reader should also
refer to the generel SDE literature such as the classic work by
\citet{kloeden1992}. See also Section~\ref{sec:higher-order-schemes}.

Note that by describing the theory for vertical transport separately, we have
implicitly assumed that the vertical motion can be treated independently of the
horizontal motion, at least within a timestep. This is usually a fair
assumption, as discussed in the next section.  However, for a more general
treatment, including iso- and diapycnal diffusivity (which leads to a
non-diagonal diffusivity tensor, $\mathbf{K}$, if the isopycnals are not
horizontal), see \citet{spivakovskaya2007}.

\subsection{Vertical timestep}
\label{sec:timestep}

Regarding the choice of timestep, it will in many cases make sense to have a far
shorter timestep for the vertical motion in an oil spill model, than for the
horizontal motion. Among the reasons for this is that available ocean data
usually have a far higher resolution in the vertical direction than in the
horizontal, and that diffusivity gradients tend to be both sharper and more
persistent in the vertical direction. Hence, inaccuracies in the vertical
transport step can lead to systematic errors in the vertical distribution of
oil, which in turn can lead to errors in, \emph{e.g.}, the prediction of
surface signature. See Section~\ref{sec:pitfalls} for some relevant examples.

\citet{visser1997} wrote down a criterion for the length of the timestep, which
is based on the requirement that the diffusivity profile should be approximately
linear over the typical length of a random step. If this criterion is satisfied,
the well-mixed condition (see Section~\ref{sec:wmc}) should be reasonably well
satisfied. He obtained
\begin{align}
    \label{eq:timesteplimit}
    \Delta t \ll \min \left| \frac{1}{K''(z)} \right|
\end{align}
where the minimum is to be taken over the entire water column, and $K''(z)$ is
the second derivative of $K(z)$ with respect to $z$. (Note that Visser did not
take the absolute value, but this is clearly an omission since $K''(z)$ can be
negative.) According to~\cite[Section 3.4]{grawe2012}, it is commonly agreed that
the timestep should be at least one order of magnitude smaller than the limit in
Eq.~\eqref{eq:timesteplimit}.

It is worth noting that if $K''(z)$ is not finite everywhere, for example
because $K(z)$ is a step function, or is a piecewise linear function with
discontinuous first derivative, then the Visser timestep condition can never be
satisfied. Fundamentally, this problem stems from the fact that the equivalence
between the advection-diffusion equation and the random walk described by
Eq.~\eqref{eq:sde}, requires both the drift and diffusion coefficients in
Eq.~\eqref{eq:sde} to be continuous. See \ref{app:equivalence} for
further details.

%TODO read Wilson (2007) and add timestep information from there

\subsection{Boundary conditions}
\label{sec:lagrangeBC}

As was discussed in Section~\ref{sec:eulerBC}, it is common in oil spill
modelling to treat the boundary at the surface differently for diffusion and
advection (advection here refers to the buoyant rise of droplets).  In a
Lagrangian model, this is straightforward to achieve by separating the advection
term and the diffusion term in Eq.~\eqref{eq:sde} into two separate steps. During
each timestep, each particle is first displaced randomly due to diffusion,
reflected from the surface or sea floor, moved upwards due to buoyancy, and
finally removed from the water column if the buoyancy brought it above the
surface.

This scheme can be formulated as the following series of operations carried out
for each particle, during each timestep, in order to update the
position, $z$. We here consider a water column of finite depth $H$ (depth
positive downwards), and a particle rising with a constant terminal speed $v_b$.

\begin{subequations}
    \begin{itemize}
        \setlength{\itemindent}{1.5em}
        \item[\bf Step 1:]
            Displace particle randomly
            \begin{align}
                z \to z + K'(z) \Delta t + \sqrt{2K(z)} \Delta W.
            \end{align}
        \item[\bf Step 2:]
            Reflect from boundaries
            \begin{align}
                z \to \left\{\begin{array}{ccc}
                    -z & \mathrm{if} & z < 0 \\
                    2H-z & \mathrm{if} & z > H\\
                    z & \mathrm{otherwise}.
                \end{array}\right.
            \end{align}
        \item[\bf Step 3:]
            Rise due to buoyancy
            \begin{align}
                z \to z - v_b \Delta t.
            \end{align}
        \item[\bf Step 4:]
            Set depth to 0 if above surface
            \begin{align}
                z \to \left\{\begin{array}{ccc}
                    0 & \mathrm{if} & z \leq 0 \\
                    z & \mathrm{otherwise}.
                \end{array}\right.
            \end{align}
    \end{itemize}
\end{subequations}

A particle that reaches the surface in Step 4 is removed from the water column
and considered ``surfaced'', corresponding to the droplet merging with a
continuous surface slick. It will then take energy in the form of breaking waves
to re-introduce surfaced oil into the water column. In that case, a fifth step
is also carried out at each timestep:

\begin{itemize}
    \setlength{\itemindent}{1.5em}
    \item[\bf Step 5:]
        If a particle is considered surfaced, it is resuspended with probability
        $p = 1 - \ue^{-\Delta t / \tau}$, in which case it is assigned random
        droplet size and depth, drawn from suitable distributions.
\end{itemize}

In Step 5, $\Delta t$ is the timestep, and the lifetime, $\tau = 1/\alpha$, is
calculated from the entrainment rate, $\alpha$ (where $\alpha$, with units
time$^{-1}$, is the decay rate of the amount of surface oil, see
Eq.~\eqref{eq:submersion_ODE}).  Note that steps 1 to 4 are applied to all
particles in the water column (\emph{i.e.}, those particles that \emph{are not}
part of the surface slick), while step 5 is applied to all particles that
\emph{are} part of the slick.

The particle scheme described by steps 1 to 5 is equivalent to Eulerian
modelling of the advection-diffusion equation with a Neumann boundary condition
at the surface, enforcing zero diffusive flux, while allowing an advective flux
\citep{nordam2019}.

%%%%%%%%%%%%%%%%%%%%%%%%%%%%%%%%%%%%%%%%
%%%%%%%%% Some Common Pitfalls %%%%%%%%%
%%%%%%%%%%%%%%%%%%%%%%%%%%%%%%%%%%%%%%%%

\section{Some examples and pitfalls}
\label{sec:pitfalls}

In this section, we describe some example calculations, and pay particular
attention to some common mistakes that should be avoided.

\subsection{Na\"i{}ve random walk}
\label{sec:naive}

\begin{figure*}[!!ht]
    \centering
    \includegraphics[width=0.8\textwidth]{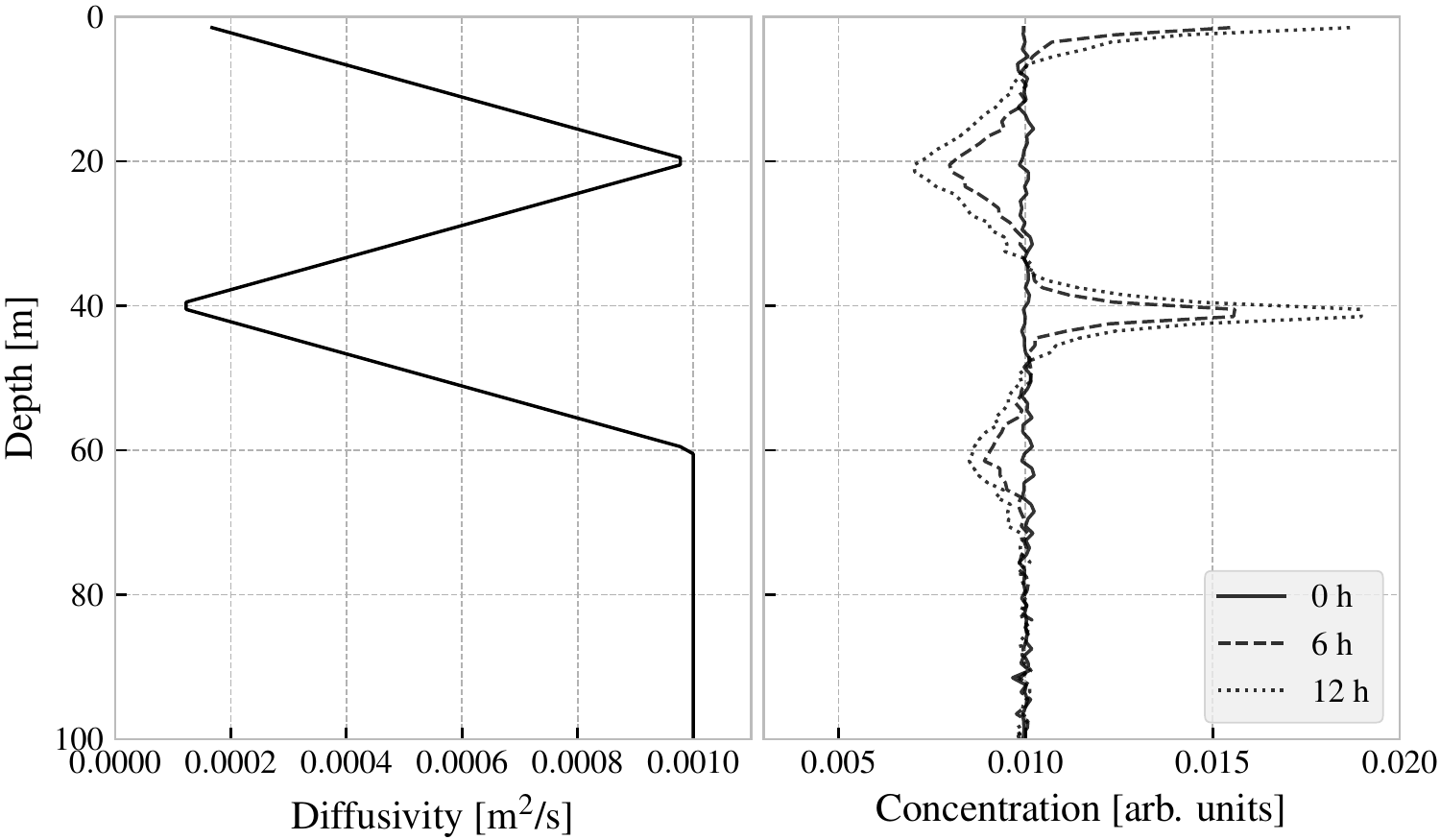}
    \caption{Concentration of initially well-mixed neutrally buoyant tracers,
    simulated with the na\"i{}ve scheme (Eq.~\eqref{eq:naive}),
        shown after different times. The number of particles was
        $N_p = \num{1000000}$, the timestep was $\Delta t = \SI{300}{\second}$,
        and concentration is calculated by bin count in 100 bins of width
        \SI{1}{\meter} each.}
    \label{fig:WMC_naive}
\end{figure*}

In the case of constant diffusivity, $K$, the random walk given by
Eq.~\eqref{eq:sde} simplifies to
\begin{align}
    \label{eq:naive_sde}
    \ud z = v_b \, \ud t + \sqrt{2K} \, \ud W(t).
\end{align}
or discretised with Euler-Maruyama
\begin{align}
    \label{eq:naive}
    z_{n+1} = z_n + v_b \Delta t + \sqrt{2K} \Delta W_n.
\end{align}
However, if $K$ is a function of position, Eq.~\eqref{eq:naive_sde} is \emph{not}
consistent with the advection-diffusion equation, and gives unphysical results
where a net transport away from regions of high diffusivity is
seen~\citep{hunter1993, holloway1994, visser1997}.  The difference between the
two schemes is the term $K'(z) \Delta t$ in Eq.~\eqref{eq:sde}, which is known
as the pseudovelocity term \citep[p.  125]{lynch2014}.

In oil spill modelling, it seems fairly common to use the random walk scheme
described by Eq.~\eqref{eq:naive}, even in combination with spatially variable
diffusivity. In, \emph{e.g.}, the plankton modelling community, the importance
of using a consistent random walk appears to have been well known for two
decades, with a particularly clear account of this issue being that of
\citet{visser1997}. In what follows, we will use the terminology of
\citeauthor{visser1997}, and refer to Eq.~\eqref{eq:naive} as the na\"i{}ve
random walk.

An investigation of this issue in the context of oil spill modelling is
presented in \citet{nordam2019naive}, where it is found that use of the
na\"i{}ve random walk scheme may lead to both over- and underprediction of the
amount of surface oil, compared to the consistent random walk scheme
(Eq.~\eqref{eq:sde}). The difference depends on the nature of the diffusivity
profile, as well as the relevant droplet size distribution.

An example is shown in Fig.~\ref{fig:WMC_naive}, where initially evenly
distributed neutrally buoyant tracers have been modelled with the na\"i{}ve
scheme (Eq.~\eqref{eq:naive}). We observe that the initially constant
concentration profile breaks down, and the tracers start to accumulate in the
regions of low diffusivity. While the example here uses neutrally buoyant
particles, it is clear that this effect can lead to errors in modelling,
\emph{e.g.}, the surfacing of small oil droplets.

%%%%%%%%%%%%%%%%%%%%%%%%%%%%%%%%%%%%%%%%%%%%%
%%%%%%%%% Step-function diffusivity %%%%%%%%%
%%%%%%%%%%%%%%%%%%%%%%%%%%%%%%%%%%%%%%%%%%%%%

\subsection{Step-function diffusivity}
\label{sec:stepfunction_example}

\begin{figure*}[!!ht]
    \begin{center}
    \includegraphics[width=0.8\textwidth]{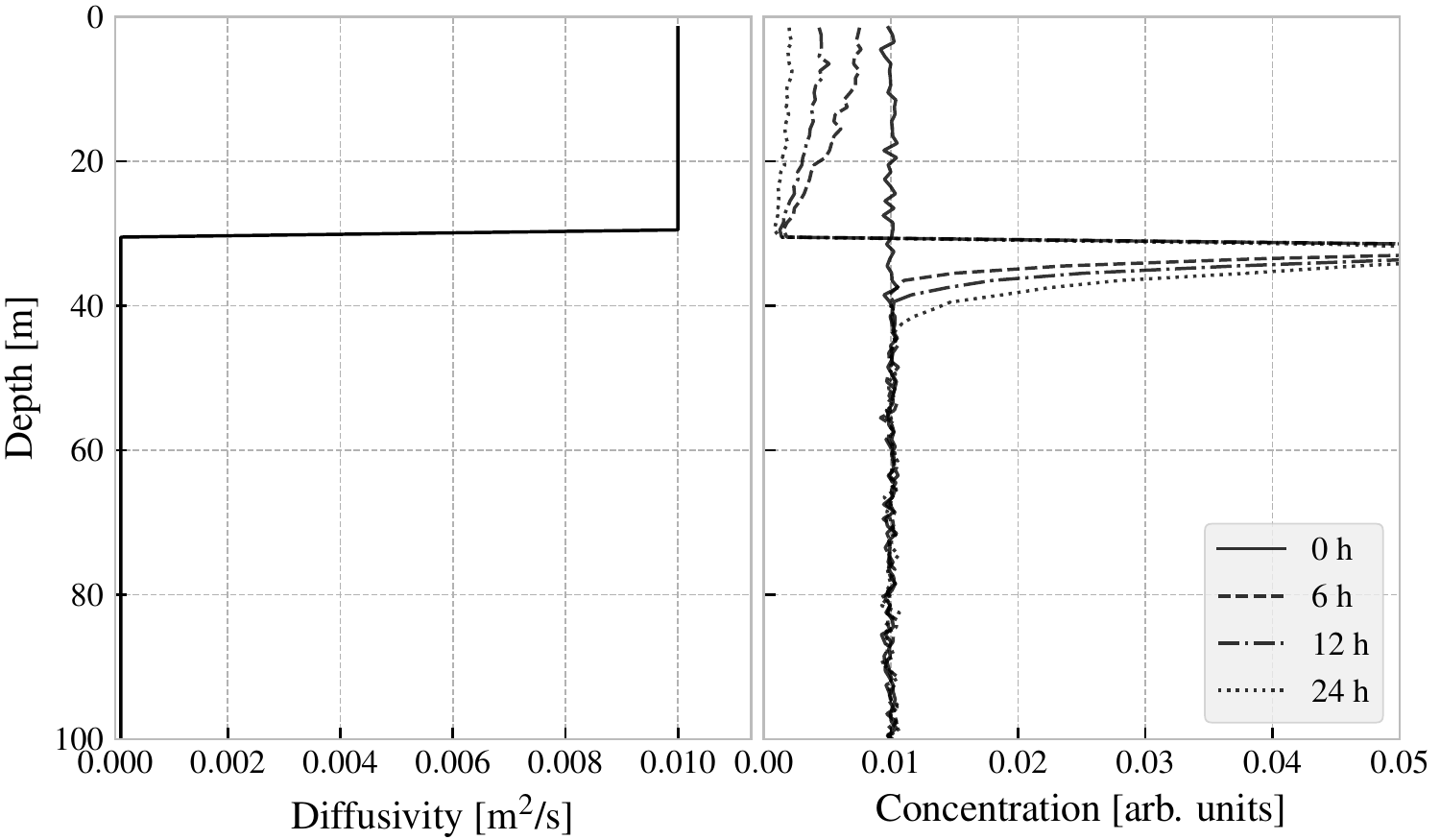}
    \end{center}
    \caption{Concentration of initially well-mixed neutrally buoyant tracers,
        shown after different times. The number of particles wa
        $N_p = \num{10000000}$, the timestep was $\Delta t = \SI{600}{\second}$,
        and concentration is calculated by bin count in 100 bins of width
        \SI{1}{\meter} each.}
    \label{fig:WMC_step_function_diffusivity}
\end{figure*}

Due to the difficulty of obtaining good data on the vertical diffusivity in
the water column, simple schemes are sometimes used. For example, a
step-function diffusivity profile may represent the well-known fact that
diffusivity tends to be higher in the mixed layer, and lower below the
pycnocline. An example of such a step-function profile used in oil spill
modelling is found, \emph{e.g.}, in \citet[Section 6.3]{dedominicis2016}:
\begin{align}
    \label{eq:K_step} K(z) = \left\{ \begin{array}{lcc}
        \SI{e-2}{\meter\squared\per\second} & \mathrm{if} & z < \SI{30}{\meter} \\
        \SI{e-4}{\meter\squared\per\second} & \mathrm{otherwise} & \\
    \end{array} \right.,
\end{align}
where depth is positive downwards. The diffusivity profile is illustrated in the
left panel of Fig.~\ref{fig:WMC_step_function_diffusivity}.

It is clear that with this diffusivity profile, the Visser timestep criterion
(Eq.~\eqref{eq:timesteplimit}) can never be satisfied, and thus we cannot expect
the well-mixed condition to be satisfied, regardless of the timestep.  The
problem can be understood intuitively by realising that particles that are in
the high-diffusivity region, but close to the transition depth to low
diffusivity, have a good probability to make a relatively large jump into the
region of low diffusivity. Once there, however, it would take this particle a
large number of steps to return to the region of high diffusivity. The net
result is that the particles tend to accumulate in the region of low
diffusivity, in violation of the WMC.

The results of a numerical test of the WMC are shown in the right panel of
Fig.~\ref{fig:WMC_step_function_diffusivity}. Neutrally buoyant particles have
been initially evenly distributed across the water column, down to a depth of $H
= \SI{100}{\meter}$. A reflecting boundary condition has been used at the
surface ($z = 0$), and at the bottom ($z = H$). Concentration profiles are shown
for different times, and it is clear that the particle count in the
high-diffusivity region is depleted, for the reason described above. As
described in the discussion of the WMC in Section~\ref{sec:wmc}, the correct
solution to the diffusion equation in this case is that the
concentration should remain constant.

While this example uses neutrally buoyant tracer particles, it is
clear that this behaviour would also be a problem in an oil spill simulation.
The effect of using this diffusivity profile is a net downwards displacement of
particles, which leads to reduced surfacing rates in an oil spill model,
particularly for small droplets with slow rise speeds.

We note that with this diffusivity profile, we have $K'(z) = 0$ everywhere,
except at $z = \SI{30}{\meter}$, where the derivative of $K$ is a Dirac
delta-function. Hence, the na\"i{}ve random walk (Eq.~\eqref{eq:naive}) and the
corrected random walk (Eq.~\eqref{eq:sde}) are identical in this case, except in
a single point, and the inclusion of the pseudovelocity term does not compensate
for the spurious downwards drift caused by the diffusivity profile.

We will now describe two approaches to avoid this problem. The first can be said
to be a ``workaround'', that modifies the diffusivity function to make it into a
smooth approximation of a step function, while the second approach uses a
different numerical scheme to solve the SDE for diffusion.

The ``workaround'' to simulating this problem would be to replace the step
function diffusivity with a smooth sigmoid function with the same asymptotic
values as the step function. In particular, the step function
\begin{align}
    \label{eq:K_step_general}
    K(z) = \left\{ \begin{array}{lcc}
        K_0 & \mathrm{if} & z < z_0 \\
        K_1 & \mathrm{otherwise}
    \end{array} \right.,
\end{align}
can be approximated as
\begin{align}
    \label{eq:K_sigmoid}
    K(z) = K_0 + \frac{K_1 - K_0}{1 + \ue^{-a(z - z_0)}},
\end{align}
where the value of the parameter $a$ determines the sharpness of the transition.
By using a diffusivity profile given by Eq.~\eqref{eq:K_sigmoid}, with large
values of $a$, true step function diffusivity can be approximated arbitrarily
well. If this is done in combination with a timestep that satisfies the Visser
criterion (Eq.~\eqref{eq:timesteplimit}), one can make sure the WMC is
satisfied. For the sigmoid diffusivity profile given by
Eq.~\eqref{eq:K_sigmoid}, the Visser timestep criterion becomes
\begin{align}
    \label{eq:timestplimit_sigmoid}
    \Delta t \ll \frac{\sqrt{3}}{18} \cdot \left| \frac{1}{a^2 (K_0 - K_1)} \right|.
\end{align}
As mentioned in Section~\ref{sec:timestep}, the timestep should be kept at least
an order of magnitude below this limit. This may make the timestep impractically
short if a large value of $a$ is chosen.

Another approach, which may be more efficient numerically, is to use the
step-function diffusivity directly, but with a different numerical scheme for
solving the SDE (Eq.~\eqref{eq:sde}). \citet{spivakovskaya2007backward} describe
an alternative to the Euler-Maruyama scheme, which they call the backward It\^o
scheme. In this scheme, the position, $z_{n+1}$, of a particle at time $t_{n+1}$
is given from its position $z_n$, at time $t_n$, by
\begin{subequations}
    \label{eq:backwardito}
\begin{align}
    \label{eq:backwardito_a}
    \tilde{z}_{n} &= z_n + \sqrt{2K(z_n)} \Delta W_n, \\
    \label{eq:backwardito_b}
    z_{n+1} &= z_n + v_b \, \Delta t + \sqrt{2K(\tilde{z}_n)} \Delta W_n,
\end{align}
\end{subequations}
where $\Delta W_n$ is \emph{the same realisation} of a Gaussian random variable
with zero mean and variance $\langle \Delta W^2_n \rangle = \Delta t$ in both
Eqs.~\eqref{eq:backwardito_a} and~\eqref{eq:backwardito_b}. Hence, the net
effect is to make a ``trial step'', to a position $\tilde{z}_n$, and then use
the diffusivity at that point, $K(\tilde{z}_n)$, in the real step. With the
backward It\^o scheme, the WMC is satisfied for a step function profile.
However, the backward It\^o scheme does not work as well as Euler-Maruyama for,
\emph{e.g.}, continuous, piecewise linear diffusivity functions. Experimentation
is encouraged, to verify that a chosen combination of diffusivity profile,
numerical scheme, and timestep satisfies the WMC to acceptable accuracy.

\subsection{Linearly interpolated diffusivity}
\label{sec:linearinterpolated_example}

\begin{figure*}[!!ht]
    \begin{center}
    \includegraphics[width=0.8\textwidth]{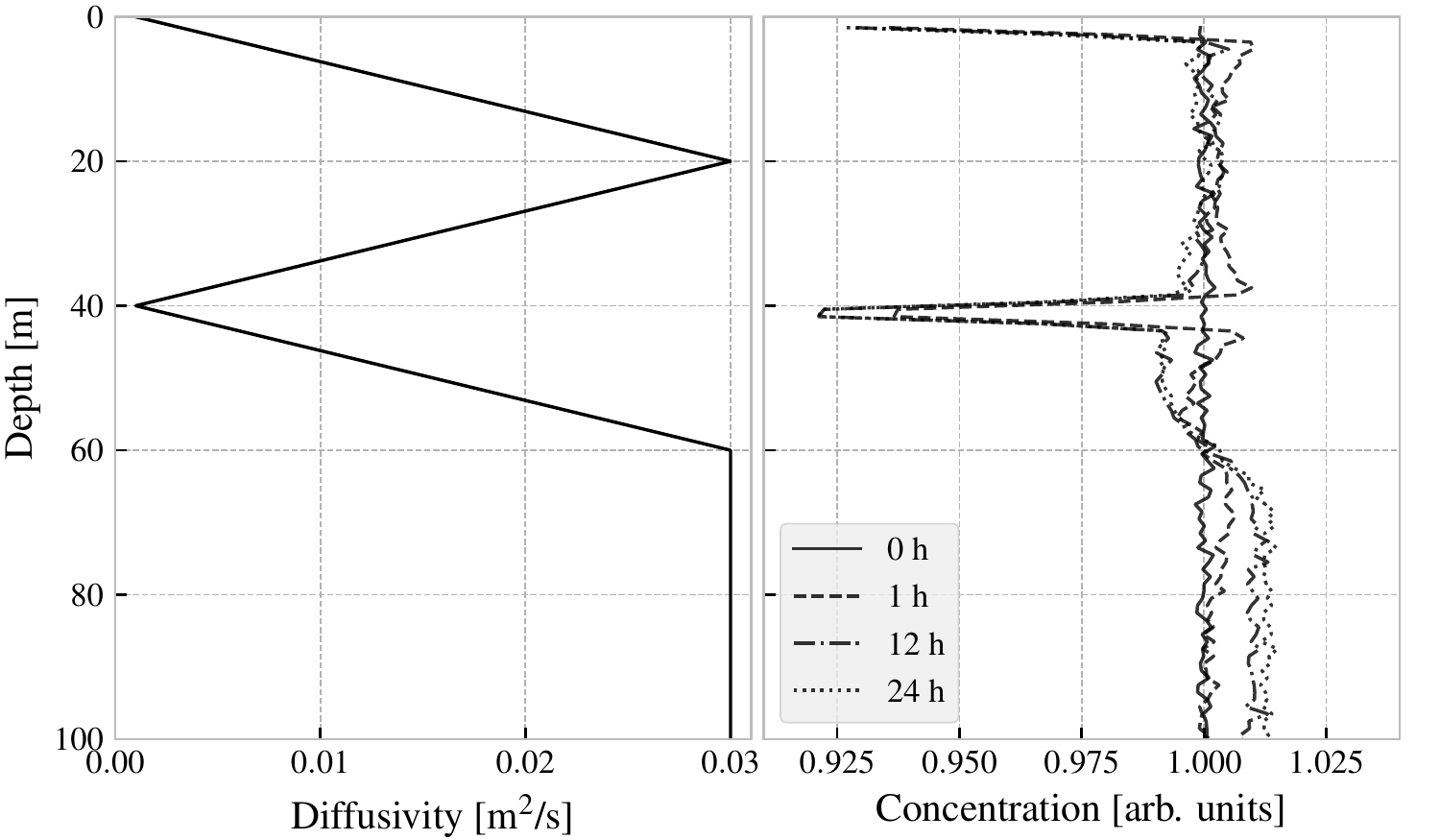}
    \end{center}
    \caption{Concentration of initially well-mixed passive tracers,
        shown after different times. The number of particles was
        $N_p = \num{1000000}$, the timestep was $\Delta t = \SI{300}{\second}$,
        and concentration is calculated by bin count in 100 bins of width
        \SI{1}{\meter} each.}
    \label{fig:WMC_linear_diffusivity}
\end{figure*}

Many ocean models provide eddy diffusivity as output, along with with current,
temperature, salinity, etc. If diffusivity is available, it can be used to drive
the random walk, but care should be taken in the interpolation of the data. In
particular, it is clear that the Visser timestep condition
(Eq.~\eqref{eq:timesteplimit}) can never be met if linear interpolation is used,
as this will give a diffusivity profile that has piecewise constant first
derivative, and hence a delta-function second derivative at each node in the
interpolation.

As an example, we have carried out a test of the WMC for a piecewise linear
diffusivity profile, as shown in the left panel of
Fig.~\ref{fig:WMC_linear_diffusivity}. While this profile is of course somewhat
artificial, it has some realistic features, in that the diffusivity goes down
towards the surface, and has a minimum at some value representing the
pycnocline (see, \emph{e.g.}, \citet{grawe2012} for a thorough discussion of the
problem of a sharp pycnocline). A passive tracer represented by $N_p =
\num{1000000}$ particles was initially evenly distributed throughout the water
column, down to a depth of \SI{100}{\meter}. Reflecting boundaries were used at
the bottom and surface.

In the right panel of Fig.~\ref{fig:WMC_linear_diffusivity}, concentration
profiles are shown for different time points. The results clearly indicate that
there are deviations from constant concentration at the minima of the
diffusivity profile, as well as below 60 m depth, where the diffusivity is
constant.  The degree to which this happens depends on the timestep, as well as
the diffusivity profile, and a sufficiently short timestep will in practical
applications remove the problem.  Nevertheless, this demonstrates that
unexpected things may happen if one uses linear interpolation of input data
without checking that the WMC is satisfied.

\subsection{Chemically dispersed oil in the mixed layer}
\label{sec:mixedlayer_example}

The final case is included as an example of a situation where a one-dimensional
oil spill model may be of practical use. We consider an idealised situation
where oil has been treated with surface dispersants, and dispersed into the
water column by means of mechanical energy, either through waves, prop wash,
water jetting or other means. The question is then, for a given droplet size,
how long may one expect the oil to remain submerged. If the oil stays submerged
for a long time, the dispersant operation may be said to have been successful.

\begin{figure*}[!ht]
    \centering
    \includegraphics[width=0.8\textwidth]{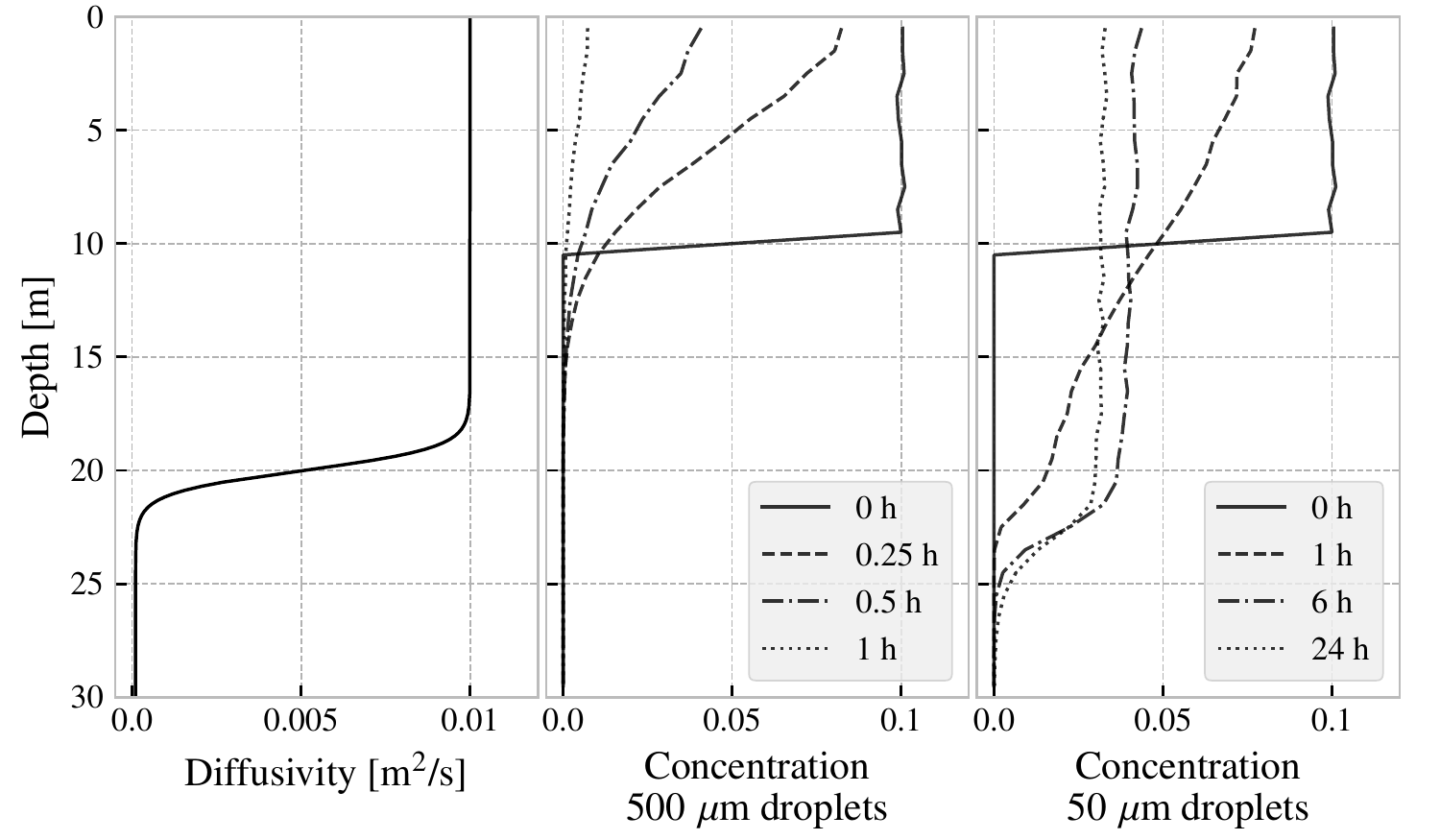}
    \caption{Left panel: Diffusivity as a function depth. Middle panel: Oil
        concentration, as a function of depth, for a droplet diameter of
        $\SI{500}{\micro\meter}$. Right panel: The same, for a droplet diameter
        of $\SI{50}{\micro\meter}$.}
    \label{fig:mixedlayer_concentration}
\end{figure*}

In this idealised case, we will consider a single droplet size, and a sigmoid
diffusivity profile giving a high diffusivity in the mixed layer, and a low
diffusivity below the pycnocline (see Section~\ref{sec:stepfunction_example}).
In particular we choose to use a diffusivity profile given by
Eq.~\eqref{eq:K_sigmoid}, with parameters 
$K_0 = \SI[per-mode=symbol]{1e-4}{\meter\squared\per\second}$,
$K_1 = \SI[per-mode=symbol]{1e-2}{\meter\squared\per\second}$,
$z_0 = \SI{20}{\meter}$, and $a = \SI{2}{\per\meter}$. The diffusivity profile
is shown in the left panel of Fig.~\ref{fig:mixedlayer_concentration}.

We consider two droplet sizes, $\SI{500}{\micro\meter}$, and
$\SI{50}{\micro\meter}$.  Assuming an oil density of
$\SI[per-mode=symbol]{0.95}{\kilogram\per\liter}$, and using
Eq.~\eqref{eq:risespeed} to calculate the rise speed, we get respectively $v_b =
\SI[per-mode=symbol]{5.4}{\milli\meter\per\second}$, and $v_b =
\SI[per-mode=symbol]{0.072}{\milli\meter\per\second}$.

Before presenting simulation results, we will try to reason about what might be
expected to happen. A useful quantity to consider here is the P\'eclet number,
\begin{align}
    \label{eq:peclet}
    \mathrm{Pe} = \frac{vH}{K},
\end{align}
which gives the ratio between advective transport, and diffusive transport. Note
that in our case, $v$ is the rise speed of the droplets, $H$ is the thickness of
the mixed layer, and $K$ is the (average) diffusivity in the mixed layer. If
$\mathrm{Pe} \gg 1$, the transport is advection-dominated (advection
here refers to the rise speed of the droplets), and if $\mathrm{Pe} \ll 1$, the
transport is diffusion-dominated. With the parameters described above, we get
$\mathrm{Pe} \approx 11$ for the large droplets, and $\mathrm{Pe} \approx 0.15$
for the small droplets.

\begin{figure*}[!ht]
    \centering
    \includegraphics[width=0.8\textwidth]{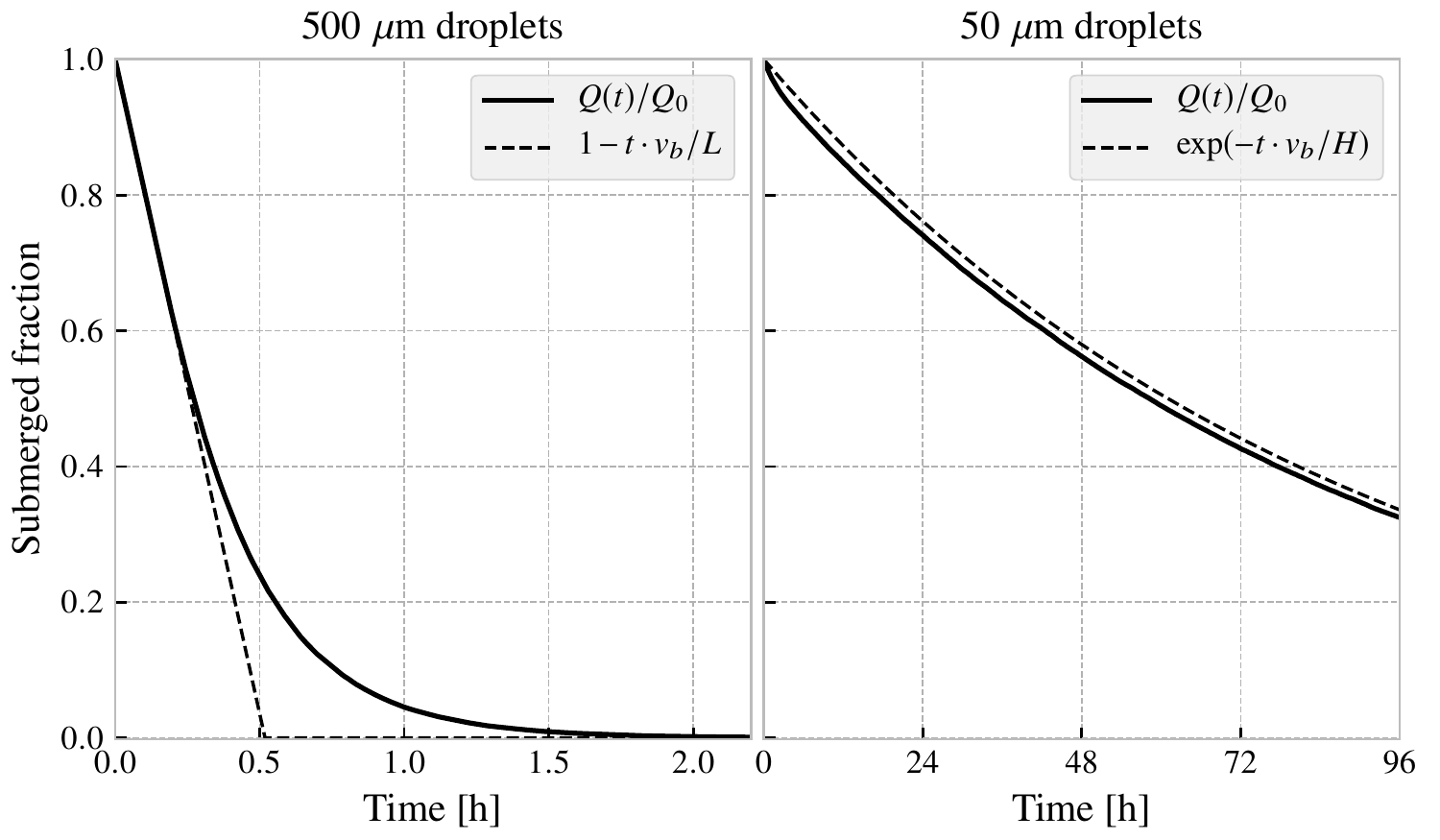}
    \caption{Left panel: Submerged fraction of oil, as a function of time, for a
        droplet diameter of $\SI{500}{\micro\meter}$. The idealised time
        development given by Eq.~\eqref{eq:submerged_large} is shown as a dashed
        line. Right panel: The same, for
        a droplet diameter of $\SI{50}{\micro\meter}$. The idealised time
        development given by Eq.~\eqref{eq:submerged_small} is shown as a dashed
        line.}
    \label{fig:mixedlayer_timeseries}
\end{figure*}

Based on these considerations, we can begin to reason about the outcome of the
dispersant operation, for the two droplet sizes we chose to look at. For the
larger droplets, the vertical transport will be dominated by the rise speed. In
the limit of zero diffusivity, the droplets will simply rise to the surface at their
terminal velocity, $v_b$. If we assume an initial amount $Q_0$ of submerged oil,
evenly distributed down to a depth $L$, then the amount of oil that remains
submerged at time $t$ is simply given by
\begin{align}
    \label{eq:submerged_large}
    Q(t) = Q_0(1 - t \frac{v_b}{L}), \;\;\;\; 0 < t < L/v_b.
\end{align}
When $t=L/v_b$, all the oil droplets have had time to reach the surface, and
there is no submerged oil remaining. While the diffusivity will never be zero in
a real case, we will see later that Eq.~\eqref{eq:submerged_large} provides a
reasonable approximation if $\mathrm{Pe} \gg 1$.

For the small droplets transport is diffusion-dominated. Hence, they will
be evenly distributed throughout the mixed layer, even if they were only
initially entrained a short distance. Furthermore, the diffusivity in the mixed
layer is sufficient to keep the remaining submerged droplets evenly distributed,
even as the surfacing begins. We conclude that the fraction of submerged droplets
that will surface during an interval $\Delta t$, is given by $v_b \Delta t / H$,
where $H$ is the thickness of the mixed layer. When a constant fraction
resurfaces during an interval, we have a first-order decay process. If the
initial amount of submerged oil is $Q_0$, then the remaining submerged oil is
given by:
\begin{align}
    \label{eq:submerged_small}
    Q(t) = Q_0 \ue^{-t/\tau}, \;\;\;\; \tau = H/v_b.
\end{align}

Thus, we find that in addition to the difference in rise speed, there is also
another difference that is relevant between advection-dominated transport (large
droplets) and diffusion-dominated transport (small droplets), and that is the
length scale. For large droplets, the entrainment depth is important, while for
small droplets, the thickness of the mixed layer is important.

We will now look at some numerical simulation results. For both droplet sizes,
we assume that the oil is initially evenly distributed down to a depth of
$L=\SI{10}{\meter}$. We run simulations using $N_p = \num{100000}$ particles.
For the diffusivity profile described above, the Visser timestep limit
(Eq.~\eqref{eq:timesteplimit}) gives $\Delta t \ll \SI{42}{\second}$, and hence
we choose $\Delta t = \SI{2}{\second}$.

In Fig.~\ref{fig:mixedlayer_concentration}, the concentration of oil droplets is
shown as a function of depth, for different times. We observe that the large
droplets rise quickly to the surface, and are not mixed any deeper than the
initial depth of $L = \SI{10}{\meter}$. For the small droplets, we observe that
they fairly quickly mix down to the pycnocline, and that the concentration
thereafter remains approximately constant with depth throughout the mixed layer.

Figure~\ref{fig:mixedlayer_timeseries} shows the remaining fraction of submerged
oil as a function of time. Additionally, the idealised time developments given
by Eqs.~\eqref{eq:submerged_large} and~\eqref{eq:submerged_small} are shown as a
dashed lines.

The purpose of this example is to illustrate some special cases that may help
provide some simple guidelines to reason about the outcome of a surface
dispersant operation. In particular, we observe that if we assume $\mathrm{Pe}
\gg 1$, then the time for the oil to surface is largely governed by the
entrainment depth and the rise velocity. On the other hand, if $\mathrm{Pe} \ll
1$, then the time development is determined by the depth of the pycnocline and
the rise velocity. The diffusivity does not appear in either case, other than in
the estimation of $\mathrm{Pe}$.

Finally, we note that it is of course not realistic to consider the entrainment
and surfacing of oil as a purely one-dimensional problem over a period of
several days, as in the right panel of Figs.~\ref{fig:mixedlayer_concentration}
and~\ref{fig:mixedlayer_timeseries}. During this time, the oil will certainly be
subject to horizontal advection and diffusion. This is of course precisely the
goal of a surface dispersant operation, and such an operation will probably be
said to be successful if the majority of the oil may be expected to remain
submerged for several days.

%%%%%%%%%%%%%%%%%%%%%%%%%%%
%%%%   Example cases   %%%%
%%%%%%%%%%%%%%%%%%%%%%%%%%%

\section{Example cases}
\label{sec:example_cases}

From the discussion in the preceding sections, it should be clear that the
vertical movement of oil in the water column is an interplay between different
effects. Entrainment moves oil from the surface, and into the water column.
Buoyancy transports oil upwards, and eventually to the surface, at a rate that
is dependent on the droplet size distribution (which in turn depends on the
conditions during entrainment).

Turbulent mixing tends to distribute the oil in the vertical. While this
diffusion process does not itself have a preferred direction, the net effect can
still be to move the center of mass of a concentration profile either up or
down, due to the reflecting boundary at the surface, vertical variation in
diffusivity, and depending on initial conditions.

In breaking wave conditions there will always be some entrainment of surface
oil. Hence, some fraction of the oil will be submerged at any time, and some
fraction will remain at the surface. The fraction at the surface will depend on
wave conditions, vertical diffusivity in the subsurface, and the state of the
oil (since droplet size distribution depends, among other things, on the
viscosity of the oil). The surface fraction will change with time even if the
wave conditions remain constant, as the oil weathers, and since the smaller
droplets may remain submerged for a very long time.

It is clear that even though oil is typically buoyant, it is quite possible for
the majority of the oil in a surface spill to be transported in the subsurface,
in a state of dynamic equilibrium between entrainment and resurfacing. As
discussed in the introduction, the vertical distribution of oil may have
significant impact on horizontal transport, due to current shear effects. The
aim of this section is to provide some examples of real oil spill scenarios
where the vertical distribution of oil is of particular importance to the
horizontal transport.

\subsection{The 1993 \emph{Braer} oil spill}
\label{sec:braer}

On January 5, 1993, MV~\emph{Braer} ran aground within 100 meters of the coast
of Shetland \citep{reed1999} during a storm. It was carrying \num{85000} tonnes
of Gullfaks crude oil, which was released into the ocean over a period of
several days \citep{spaulding1994}. During the event, model forecasts were made
available, but failed to accurately predict the movement of the oil
\citep{turrell1994, turrell1995}. Later, several hindcast modelling studies were
made (see, \emph{e.g.}, \citet{spaulding1994, turrell1994, proctor1994}).

While the wind was mainly flowing towards the north-east, much of the oil moved towards
the south, with oil found in the sediments up to \SI{100}{\kilo\meter} to the
south of the spill site \citep{proctor1994}. One explanation would be 
that the relatively light Gullfaks crude dispersed as small droplets in the
strong winds present during the spill, causing a large fraction of the spilled
oil to be transported towards the south by the subsurface currents. An
additional relevant mechanism is that of oil-mineral aggregation (which has not
been discussed in this chapter), which may cause oil to sink when associated
with high density mineral particles.

\subsection{The 2011 \emph{Golden Trader} oil spill}
\label{sec:goldentrader}

On September 10, the bulk carrier MV \emph{Golden Trader} collided with a
fishing vessel off the north-west coast of Denmark. There were no casualties,
but some bunker fuel was spilled from MV \emph{Golden Trader}. The amount was
later estimated at 150 tonnes. During the first two days after the spill,
approximately 50 tonnes of oil were collected by Danish response vessels. After
this, the wind picked up, and no further observations of oil were reported until
September 15, when oil reached the Swedish shore. On September 16, it became
clear that a significant amount of oil (estimated amount 25--30 tonnes) had
beached \citep{marinesafety2012}.

The distance from the release point to the site of the beaching is more than
\SI{250}{\kilo\meter}. Only approximately \SI{15}{\kilo\meter} of the shoreline
was heavily oiled \citep{itopf2011}. Combined with the fact that beaching appears
to have occurred over a period of half a day or more, this indicates that the
slick may have been elongated in the wind direction, and relatively narrow in
the cross-wind direction, as discussed by, \emph{e.g.}, \citet{johansen1982} and
\citet{elliott_shear_1986}.

To the best of our knowledge, no detailed hindcast of this incident has been
published. Such a hindcast would however be an interesting exercise. It seems
likely that a number of model processes will impact the arrival time of the
oil and the site of the beaching, including droplet size distribution, vertical
mixing, and possibly Stokes' drift \citep{brostrom2014}.

%%%%%%%%%%%%%%%%%%%%%%%%%%%%%%%%%%%%%%%%%%%%
%%%%%%%% Advanced numerical topics %%%%%%%%%
%%%%%%%%%%%%%%%%%%%%%%%%%%%%%%%%%%%%%%%%%%%%

\section{Advanced topics and further reading}

Historically, the mathematical and technical details of Lagrangian particle
schemes have received limited attention in papers on oil spill modelling (see,
\emph{e.g.}, \citet{nordam2019naive} and references therein). A challenge is that
the mathematical literature on Stochastic differential equations is often very
technical, and not very accessible to non-specialists. However, there exists a
large body of work on the modelling of plankton, fish eggs, sediment particles,
atmospheric dispersion, etc., where these schemes are treated more rigorously
that what is commonly seen in the oil spill modelling literature. Much of this
work is formulated in terms of familiar concepts from applied oceanography, and
may be more or less directly applied to the transport part of oil spill
modelling.

In this section, we discuss some advanced topics, and recommend some further
reading for those who are interested in the details of these topics.

\subsection{Higher-order SDE solvers}
\label{sec:higher-order-schemes}

Earlier, we used the Euler-Maruyama scheme to discretise Eq.~\eqref{eq:sde},
obtaining the following iterative scheme for particle positions:
\begin{align}
    z_{n+1} = z_n + \big(w + K'(z_n)\big) \Delta t + \sqrt{2K(z_n)} \Delta W_n.
    \nonumber
\end{align}
However, just like the Euler scheme is the simplest, and least accurate, ODE
solver, so the Euler-Maruyama scheme is the simplest and least accurate SDE
solver. Switching to higher-order schemes should in principle give improved
accuracy at the same timestep, or reduce computational effort by allowing a
longer timestep to be used.

For SDE schemes, two types of convergence exist, weak and strong. Convergence
in the weak sense means that for a large number of particles, the distribution
of particles will converge towards the \emph{true} distribution (which may or
may not be known) as the timestep goes to zero. Technically, weak convergence
is expressed in the following way: If, for a numerical SDE scheme, and for
sufficiently short timesteps $\Delta t$, there exists a constant $C$, such that
\begin{align}
    \label{eq:weak}
    |\langle f(z_N) \rangle - \langle f(z(t_N)) \rangle | < C \Delta t^\gamma,
\end{align}
then the scheme is said to have order of convergence $\gamma$ in the weak
sense. Here, $z_N$ is the numerical approximation at time $t_N$, and $z(t_N)$
is the true solution at the same time, and the angle brackets indicate ensemble
average over many independent particles. The functions $f$ are continuous
functions that have polynomial growth, and are at least $2(\gamma+1)$ times
differentiable. Since this class of functions include all the integer powers of
$z$, it follows that the moments of the distribution converge if the scheme
converges in the weak sense. As any distribution is uniquely defined by its
moments, this means that the modelled distribution converges to the true
distribution.

Convergence in the strong sense is also called pathwise convergence. If, for a
numerical SDE scheme, and for sufficiently short $\Delta t$, there exists a
constant $C$, such that
\begin{align}
    \label{eq:strong}
    \langle | z_N - z(t_N) | \rangle  < C \Delta t^\gamma,
\end{align}
then the scheme is said to have order of convergence $\gamma$ in the strong
sense.

The Euler-Maruyama scheme has orders of convergence $1/2$ in the strong sense,
and $1$ in the weak sense.  Higher-order schemes exist, but the complexity of
the schemes grows fast as the order increases. An example of a higher order
scheme is the 1st-order Milstein scheme, which has order of convergence 1, in
both the strong and the weak sense (see, \emph{e.g.},
\citet[p.~345]{kloeden1992}). Applied to our SDE for advection-diffusion
problems (Eq.~\eqref{eq:sde}), the 1st-order Milstein scheme yields
\begin{align}
    \label{eq:milstein}
    z_{n+1} = z_n
    + \big(w + K'(z_n)\big) \Delta t
    + \sqrt{2K(z_n)} \Delta W_n
    + \frac{1}{2} K'(z_n) (\Delta t - \Delta W_n^2).
\end{align}

\citet{grawe2012} argue that in some cases, the Euler-Maruyama scheme is simply
inadequate, even with very short timesteps. The example they give is that of a
strong, sharp pycnocline where the diffusivity will drop almost to zero at the
steepest point of the density gradient. In such a case, passive tracers should
cross the pycnocline very slowly, a behaviour that is modelled far more
accurately by the 1st-order Milstein scheme, due to its higher order of
convergence in the strong sense.

For a clear and readable presentation of a range of numerical SDE schemes, with
a view to marine particle tracking applications, the interested reader is
referred to \citet{grawe2011, grawe2012}. Note however that some of the schemes
have been found to contain small mistakes, hence it is also advisable to consult
other sources prior to implementation, for example the classic work by
\citet{kloeden1992}.

\subsection{Autocorrelated velocity or acceleration}
\label{sec:ARn}

Implementing a random walk scheme that makes random displacements at each
timestep, with no correlation in time, makes the implicit assumption that a
moving particle can instantly change its velocity. This may seem unreasonable.
Furthermore, when very short timesteps are used we find that particle speed
becomes arbitrarily large, since the average step-length is proportional to
$\sqrt{2K\Delta t}$, and we have
\begin{align}
    \label{eq:speedlimit}
    \lim_{\Delta t \to 0} \frac{\sqrt{2K \Delta t}}{\Delta t} = \infty
\end{align}
for any positive $K$. Note, however, that while these points sound unreasonable
from a physical point of view, there is no problem in using the random walk
scheme with short timesteps. Eq.~\eqref{eq:sde} was derived to be consistent
with the advection-diffusion equation, and in the limit $\Delta t \to 0$, $N_p
\to \infty$, the distribution of particles \emph{will} converge to correct
distribution, almost surely\footnote{The term ``almost surely'' is used in the
technical sense meaning ``with probability 1''.} (provided $K(z)$ and $K'(z)$
are sufficiently smooth functions).

In fact, the ``infinite speed'' of the particle is a feature which is built into
the model from the start: The Wiener process, $W(t)$, whose increments appear in
Eq.~\eqref{eq:sde}, has infinite total variation on any interval of non-zero
length \citep[pp.~157--158]{brzezniak1999}. The apparent problem stems only from
trying to extract a physically meaningful ``speed'' from a model that does not
contain the speed of the particle as a variable.

Nevertheless, it might in some cases be desirable to have a more physically
realistic random walk model. Recall that what we have been calling diffusivity,
is in reality a parameterisation of mixing due to turbulence. If we consider a
neutral tracer in a field of turbulent eddies, it is clear that the velocity at
one instant will be at least somewhat correlated to the velocity a short time
later. This behaviour can also be captured in numerical modelling.

\cite{lynch2014} describe a hierarchy of random walk models with different
degrees of autocorrelation. The standard random walk that we have been
considering so far is called AR0 in this hierarchy, as it has no autocorrelation
in the displacement at each step. (Note that the position of a particle does of
course have autocorrelation, as the position at time $t_n$ depends on the
position at time $t_{n-1}$.)

The next level of the hierarchy is called AR1, where the displacement at each
step is related to the displacement at the previous step. In this scheme, there
is autocorrelation not only in the position of a particle, but also in its
velocity. This is in a way a more realistic model, as in reality, the movement
of a particle from one instant to the next is correlated, with the
de-correlation time being dependent on the turbulent fluctuations.

The original Langevin equation was formulated to describe Brownian motion,
\emph{i.e.}, the apparently random motion of small particles in fluids, caused
by collisions with the molecules of the fluid \citep{langevin1908, lemons1997},
and reads
\begin{align}
    \label{eq:langevin}
    m \frac{\ud^2 x}{\ud t^2} = - 6 \pi \mu a \frac{\ud x}{\ud t} + X.
\end{align}
Here, $m$ is the mass of the particle, $- 6\pi \mu a \frac{\ud x }{ \ud t}$ is
the drag force from the bulk fluid, and $X$ is a random force representing the
collisions of molecules. Hence, this is simply Newton's second law, with a
random component in the force.

Along the same lines, \citet{lynch2014} write down a general equation for an AR1
scheme as,
\begin{align}
    \label{eq:AR1}
    \frac{\ud^2 z}{\ud t^2} + \frac{1}{\tau}\frac{\ud z}{\ud t} = \eta,
\end{align}
where $\eta$ is some random process acting as a forcing, and $\tau$ is a
timescale for decay of the velocity if no forcing is applied. Written as a pair
of coupled first-order SDEs in standard notation, this becomes
%
% TODO Check that this is correct (there may be a missing prefactor in the noise term)
\begin{align}
\begin{aligned}
    \label{eq:AR1-sde}
    \ud v &= -v \frac{1}{\tau} \, \ud t + \eta\, \ud W_t, \\
    \ud z &= v \, \ud t.
\end{aligned}
\end{align}
In this model,
there is a time-correlation in the movement of the particle, since the velocity
will only change by a small amount between timesteps. This is also
called a ``random flight''. It should be noted that an AR1 scheme is
fundamentally different from an AR0 scheme, in that it is \emph{not} consistent
with the diffusion equation. And that is of course the argument for using this
schemes in the first place, since what we are trying to model is turbulent
mixing, and not pure diffusion.

% TOR: tar du en titt på wilson2007 for å sjekke at jeg har forstått
% konklusjonen riktig?
AR1 schemes have a long history of usage in dispersion models for the
atmosphere~\citep{thomson1987}. However, comparisons between this and the AR0
scheme have shown small differences in the far-field~\citep{wilson2007critical}.
This implies that differences in the results of an oil spill simulation are
unlikely to be substantial.

A review by \cite{spaulding2017} mentions regarding an AR1 scheme that ``[u]se
of this higher order model is possible if one has accurate estimates of the
currents and dispersion'', but does not elaborate further. A recent paper by
\citet{cui2018oil} solves the so-called Maxey-Riley equation, describing the
inertia and drag forces on individual oil droplets due to turbulent motion of the
surrounding waters, and compare the results to a regular random walk scheme.
However, this work considers only small spatial scales underneath breaking
waves, and the effects on larger scales are not investigated.

In conclusion, AR1 (or even higher order) schemes do not appear to be commonly
used in oil spill modelling. In addition to the increased mathematical and
numerical complexity, a practical problem in using an AR1 scheme is that one can
no longer use the eddy diffusivity directly, but must instead obtain estimates
of the parameters $\tau$ and $\eta$ in Eq.~\eqref{eq:AR1}. The interested reader
is referred to \citet{lynch2014}, and references therein, as well as the
literature on atmospheric dispersion (see, \emph{e.g.}, \citet{thomson1987},
\citet{wilson1993}).
%TODO add more references

\subsection{Reconstructing a concentration field from particles}
\label{sec:concentration}

As discussed in Section~\ref{sec:lagrangian}, our random walk scheme is in some
sense equivalent to the advection-diffusion equation. The link is that each
particle, at time $t_n$, represents a \emph{sample} from the distribution at
that time, where the distribution develops according to the advection-diffusion
equation. If we want to (approximately) reconstruct the distribution from the
particles, there are several different approaches, and which is most suitable
may depend on the application. We discuss these in one dimension in this
section, but generalisation to several dimensions is natural.

The simplest approach is the so-called box count or bin count, which consists of
dividing the region of interest into discrete bins, and counting the number of
particles in each bin. The concentration in each bin is then proportional to
that number, weighted by the particle mass if each particles represents a
different mass. This is exactly the same as a weighted histogram of particle
positions.

We let our cells have constant size $\Delta z$, and define cell $i$ by
$(i - 1) \Delta z \leq z < i \Delta z$. Furthermore, let particle $j$ have
position $z_j$ and represent a mass $m_j$. Then the concentration, $C_i$,
in cell $i$, is given by
\begin{align}
    \label{eq:Ci}
    C_i = \frac{1}{\Delta z} \sum m_j \textrm{ for all } j
    \textrm{ such that } (i-1) \Delta z \leq z_j < i \Delta z,
\end{align}
where $N_p$ is the total number of particles. A natural question to ask is then
how the error in the concentration scales with the number of particles and the
cell size.

We recall that the particle positions are essentially random samples,
and two simulations will in general give somewhat different concentration fields
due to this randomness. The difference between the \emph{true} distribution
(which is usually unknown), and the reconstructed distribution based on $N_p$
samples, is called the sampling error. One may see from the Central Limit
Theorem \citep[p.~308]{billingsley1979}, that the sampling error scales as
$1/\sqrt{N_p}$ where $N_p$ is the number of \emph{independent} samples. This
means that increasing the number of particles by a factor of 10 will only reduce
the error by a factor of $\sqrt{10}$.

Regarding cell size, there is a choice to be made between resolution
and sampling error. In the case where the entire domain is covered by just one
cell, then all the particles will be inside that cell, which is of course
correct, but also a useless result. On the other hand, if there are so many
cells that most cells have either 0 of 1 particles, then the result is
completely dominated by random sampling noise. The challenge is to use enough
cells to resolve those changes in concentration that are of interest, and enough
particles to give a reasonably smooth result.

Box counting often leads to very noisy concentration fields. In particular, if
one is interested in the most dilute concentrations, the results are guaranteed
to be noisy, because the most dilute concentrations are by definition
represented by only a small number of particles. However, even the higher
concentrations may be noisy. A common way to tackle this problem is to use a
kernel, where each particle is not treated as a point, but as a distribution
with a finite extent. In statistics, this is called Kernel Density Estimation
(KDE). For further details, see, \emph{e.g.}, \citet{silverman1986}.

The kernel function, $\kappa(z)$ must be a positive
function, with the property
\begin{align}
    \label{eq:normalised_kernel}
    \int_{-\infty}^{\infty} \kappa(z) \, \ud z = 1.
\end{align}
Usually, $\kappa(z)$ is also symmetric, and with a maximum at $z=0$. Then the
concentration field, $C(z)$, is given by
\begin{align}
    \label{eq:kde}
    C(z) = \frac{1}{N_p} \sum_{j=0}^{N_p} \frac{m_j}{\lambda j} \, \kappa \left( \frac{z -
    z_j}{\lambda_j} \right),
\end{align}
where $z_j$ and $m_j$ are as before the position and mass of particle $j$, and
$\lambda_j$ is called the bandwidth of particle $j$. For a given kernel
function, increasing the bandwidth will widen the kernel, and give a smoother
(but less detailed) concentration field. Hence, the choice of both kernel and
bandwidth becomes important, with the bandwidth typically more important than
the kernel (for standard choices of kernel function).
% TODO add specific references for choice of bandwidth

\begin{figure*}
    \centering
    \includegraphics[width=0.8\textwidth]{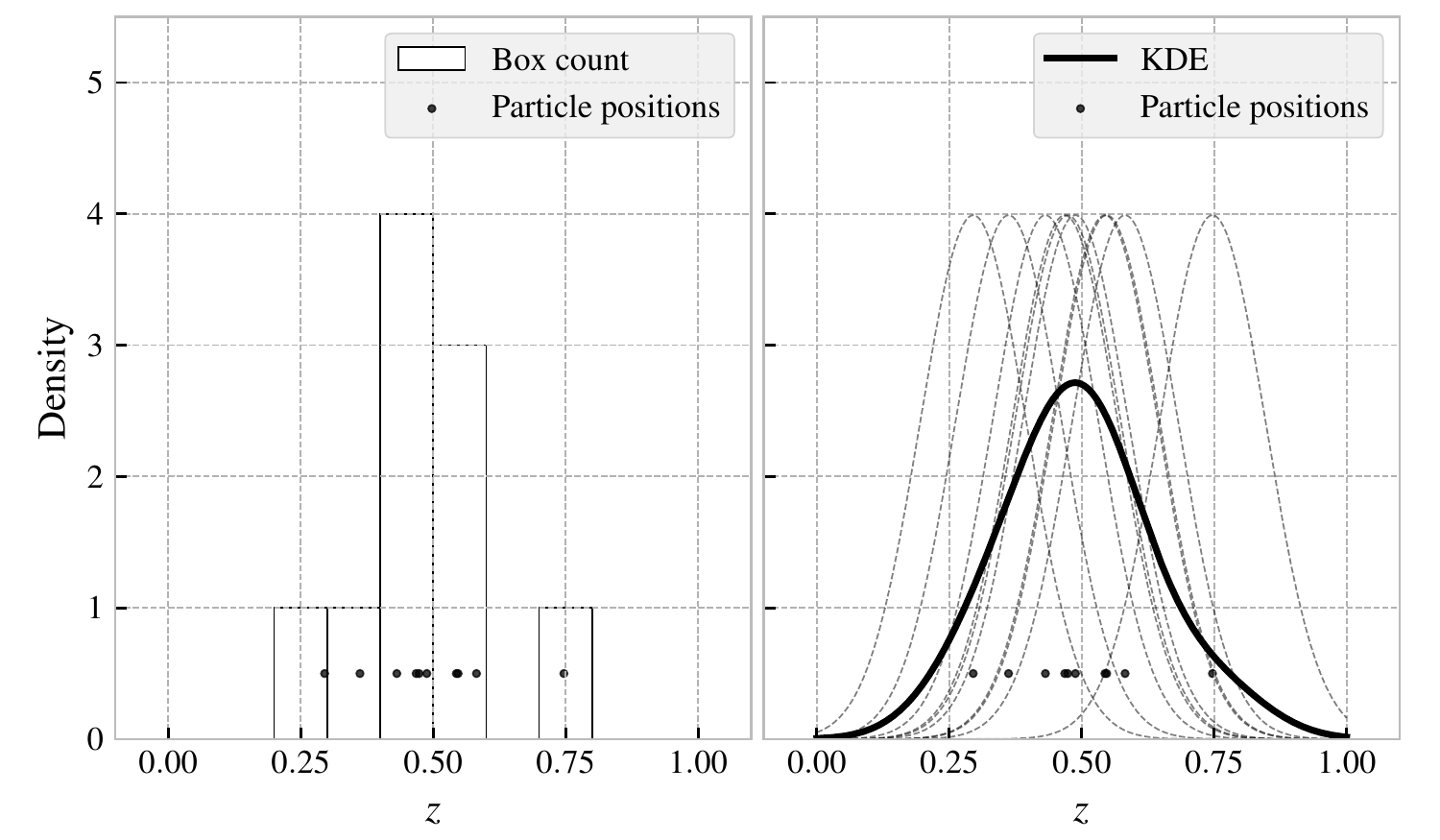}
    \caption{Probability density reconstruction based on $N_p = 10$ random particle
        positions, drawn from a normal distribution. In the left panel, box
        count (histogram) is used. In the right panel, KDE is used with a
        Gaussian kernel (unit variance) and a bandwidth of $\lambda = 0.1$. The
        bandwidth-scaled kernel of each individual particle is shown as a thin,
        dashed line. In both cases, the particle positions were the same.}
    \label{fig:KDE_comparison}
\end{figure*}

An example comparing box count and KDE is shown in
Fig.~\ref{fig:KDE_comparison}. Here, 10 random particle positions were drawn
from a Gaussian distribution with mean 0.5, and standard deviation 0.1. In the
left panel, a box count on 10 cells of length $\Delta z = 0.1$ was used. In the
right panel, KDE was used, with a Gaussian kernel (unit variance), and a
constant bandwidth of $\lambda = 0.1$. It is clear that the density
reconstructed by KDE gives a much smoother result, and a better approximation of
the underlying distribution.

For further reading on this topic, see \emph{e.g.}, \citet[Chapter
8]{lynch2014}.

%%%%%%%%%%%%%%%%%%%%%
%%%%   Summary   %%%%
%%%%%%%%%%%%%%%%%%%%%

\section{Summary}
\label{sec:summary}

The aim of this chapter has been to introduce the reader to most processes that
are relevant for modelling of the vertical distribution of oil spilled at sea,
with the exception of near-field plume modelling. It is our hope that the reader
will find themselves able to understand, and indeed implement, numerical models
for the relevant vertical transport processes. We have also tried to give
references to further reading, indicate what some uncertainties are, and point
out some examples of problems that require more research.

An eternal problem of oil spill modelling is that of input data. We know that
there are always large uncertainties in meteorological input, perhaps most
importantly in the currents. Likewise, modelling vertical eddy diffusivity is a
research field in itself, and it may be difficult to know what diffusivity
profiles to use as input to the oil spill modelling.

It is worth remembering that oil spill modelling is, to some degree, an exercise
in pragmatism. This is especially true for operational modelling in support of
oil spill response, where time is of the essence, and good data might be hard to
obtain. For model development in general, it is also worth considering where the
largest uncertainties lie, and putting the effort there.

In general, we encourage experimentation and testing, to make sure models
satisfy those exact solutions that are known to exist, such as the WMC. We
would also suggest that some attention is paid to numerical schemes, even though
these are usually not the source of the largest errors. Using bad numerical
schemes can lead to large and systematic errors, and can mask the improvements
of model development in other areas.

We have also made a point of introducing, \emph{e.g.}, the P\'eclet number and
the Richardson number, that can sometimes be used to characterise situations as
either diffusion dominated, or advection dominated. It is easy to think that the
job of an oil spill modeller is to run an oil spill model on a set of input
data, but by taking a critical look at those data one can sometimes reason quite
successfully about the expected outcome of a situation.

Finally, we would like to encourage our readers to stay curious, to experiment
with models, to read papers from related research fields, and to contribute
to the literature by publishing detailed descriptions of new modelling
developments.

%%%%%%%%%%%%%%%%%%%%%%%%%%%
%%%%%%% Appendices %%%%%%%%
%%%%%%%%%%%%%%%%%%%%%%%%%%%

\appendix
\section{Equivalence between Eulerian and Lagrangian pictures}
\label{app:equivalence}

The development of a concentration field under transport and mixing may be
described by the Partial Differential Equation (PDE) known as the
advection-diffusion equation. This is called an Eulerian approach, and is
characterised by an equation that describes how the concentration at fixed
locations changes in time. The same process may also be described by an
ensemble of ``particles'' which experience directed motion due to advection,
and random motion due to diffusion. This approach is called Lagrangian, and
is characterised by an equation that describes how the position of a
particle changes with time.

The link between the Eulerian and the Lagrangian picture is that the
concentration field described in the Eulerian picture, if normalised,
describes a probability distribution for where the Lagrangian particles will
be found. Conversely, calculating the position of a Lagrangian particle is
the same as drawing a sample from the probability distribution, and with a
large number of samples, the distribution can be reconstructed
approximately. We will here demonstrate how to obtain a random walk which is
equivalent to the advection diffusion equation.

Consider a diffusion process described by the general Stochastic Differential
Equation (SDE)
\begin{align}
    \label{eq:diffusionprocess}
    \ud z = a(z, t) \, \ud t + b(z, t) \, \ud W(t),
\end{align}
where $a(z, t)$ and $b(z,t)$ are ``moderately smooth functions'' \citep[p.
37]{kloeden1992}, and $\ud W (t)$ are the increments of a standard Wiener
process \citep[p. 40]{kloeden1992}.  Further conditions also apply, though
these may be less important in practice. For details see, \emph{e.g.},
\citet[pp. 96--102]{gihman1972}.

For this diffusion process, the Fokker-Planck equation (also known as the
Kolmogorov Forward equation) for evolution of the transition probability
density, $p(z_0, t_0, z, t)$, from an initial position $z_0$ at time $t_0$,
to a position $z$ at a later time $t$, is \citep[p. 37]{kloeden1992}:
\begin{align}
\begin{aligned}
    \label{eq:fokkerplanck}
    \frac{\p p(z_0,t_0,z,t)}{\p t} & =
    \frac{1}{2} \frac{\p^2}{\p z^2}\Big( b^2(z,t) \, p(z_0,t_0,z,t) \Big) \\
    & - \frac{\p}{\p z} \Big( a(z, t) \, p(z_0,t_0,z,t)\Big).
\end{aligned}
\end{align}
We observe that the Fokker-Planck equation is a PDE, and like the
advection-diffusion equation, it describes the time-development of a
distribution. For a particle initially at position $z_0$ at $t_0$,
undergoing the random motion described by Eq.~\ref{eq:diffusionprocess},
the probability density function for the position, $z$, at a later time,
$t$, may be obtained by the Fokker-Planck equation. If we consider
instead a large ensemble of particles, all starting out at $z_0$ at
$t_0$, then at a later time $t$, they will be distributed according to
$p(z_0, t_0, z, t)$, with many particles in areas of high probability
and few particles in areas of low probability.

This is equivalent to the evolution of a concentration field from an
instantaneous point source, as described by the advection-diffusion
equation. Hence, our goal is to obtain the SDE which has the
advection-diffusion equation as its Fokker-Planck equation. Then we know
that the distribution of an ensemble of particles will develop according to
the advection-diffusion equation, and thus we can use the distribution of
particles to approximately reconstruct the concentration field.

Going back to Eq.~\eqref{eq:fokkerplanck}, we drop the arguments to $a$, $b$ and
$p$ for brevity, and rewrite the equation a bit, and we get
\begin{align}
    \label{eq:fokkerplanck2}
    \frac{\p p}{\p t} &=
    \frac{1}{2} \frac{\p}{\p z} \left( b^2 \frac{\p p}{\p z} \right)
    - \frac{\p}{\p z}
    \left[ \left( a - \frac{1}{2} \frac{\p b^2}{\p z} \right) p \right].
\end{align}
We then compare Eq.~\eqref{eq:fokkerplanck2} to the advection-diffusion
equation, with advection $w(z,t)$ and diffusion $K(z,t)$:
\begin{align}
    \label{eq:ad2}
    \frac{\p C}{\p t}
    = \frac{\p}{\p z} \left( K \frac{\p C}{\p z} \right)
    - \frac{\p }{\p z}\left( w C \right).
\end{align}
By demanding that $C$ should be proportional to $p$ at all times, we find that
each term in Eq.~\eqref{eq:fokkerplanck2} must be equal to the corresponding
term in Eq.~\eqref{eq:ad2}. We thus obtain
\begin{subequations}
\begin{align}
    \label{eq:equivalent_terms}
    K = \frac{b^2}{2} \; &\Rightarrow \; b = \sqrt{2K} \\
    w = a - \frac{1}{2} \frac{\p b^2}{\p z} \; &\Rightarrow \; a = w + \p_z K
\end{align}
\end{subequations}
Hence, the SDE whose probability density is described by the advection-diffusion
equation is
\begin{align}
    \ud z = (w + K'(z))\, \ud t + \sqrt{2 K(z)} \, \ud W,
\end{align}
where $K'(z) = \p_z K$.

Note that since both $a(z, t)$ and $b(z, t)$ in Eq.~\eqref{eq:diffusionprocess} must
be continuous, we find that both $K(z)$ and $\partial_z K (z)$ must be
continuous for the conditions mentioned above to be satisfied. Hence, the
equivalence with the advection-diffusion does not hold for, \emph{e.g.},
step-function diffusivity, or piecewise linear diffusivity profiles with
discontinuous first derivatives, such as a linearly interpolated profile.

\bibliographystyle{humannat} \bibliography{vertical_transport}

\end{document}